\documentclass[aip, pop, reprint, onecolumn]{revtex4-1}

\usepackage{natbib}
\usepackage{subfigure}
\usepackage{graphicx}
\usepackage{epstopdf}
\usepackage{amsmath}

\usepackage{amssymb}
\usepackage{color}
\usepackage{xspace}

\newcommand{\sgn}{\operatorname{sgn}}

\newcommand\etc{\textit{etc.}\xspace}
\newcommand\eg{\textit{e.g.}\xspace}
\newcommand\ie{\textit{i.e.}\xspace}

\renewcommand\Re{\mbox{$\mathrm{Re}$}}
\renewcommand\Im{\mbox{$\mathrm{Im}$}}

\newcommand{\field}[1]{\mathbb{#1}}
\newcommand{\vpar}{\ensuremath{v_{\parallel}}}
\newcommand{\vperp}{\ensuremath{v_{\perp}}}
\newcommand{\xpar}{\ensuremath{x_{\parallel}}}
\newcommand{\xperp}{\ensuremath{x_{\perp}}}

\providecommand\bnabla{\boldsymbol{\nabla}}
\providecommand\bkappa{\boldsymbol{\kappa}}

\providecommand\bnabla{\boldsymbol{\nabla}}

\def \kpar {\mbox{$k_{\parallel}$}}
\def \Lpar {\mbox{$L_{\parallel}$}}
\def \vth {\mbox{$v_{\mathrm{T}}$}}

\def \bc {\mbox{$b_{\mathrm{m}}$}}
\def \kyc {\mbox{$k_{y\mathrm{c}}$}}

\def \O {\mbox{$\Omega$}}
\def \os {\mbox{$\omega_*$}}
\def \osT {\mbox{$\omega_*^{\mathrm{T}}$}}
\def \OsT {\mbox{$\Omega_*^{\mathrm{T}}$}}
\def \ost {\mbox{$\tilde{\omega}_*$}}
\def \Os {\mbox{$\Omega_*$}}
\def \ot {\mbox{$\omega_t$}}
\def \op {\mbox{$\omega_{\parallel}$}}
\def \od {\mbox{$\omega_d$}}

\def \odt {\mbox{$\tilde{\omega}_d$}}
\def \Odt {\mbox{$\tilde{\Omega}_d$}}
\def \odo {\mbox{$\omega_{d0}$}}

\def \Reff {\mbox{$R_{\mathrm{eff}}$}}
\def \qeff {\mbox{$q_{\mathrm{eff}}$}}

\begin{document}

\title{Collisionless microinstabilities in stellarators III - the ion-temperature-gradient mode}

\author{G. G. Plunk}
\email{gplunk@ipp.mpg.de}
\affiliation{Max Planck Institute for Plasma Physics, EURATOM Association, Wendelsteinstr. 1, 17491 Greifswald, Germany}
\affiliation{Max-Planck/Princeton Research Center for Plasma Physics}
\author{P. Helander}
\affiliation{Max Planck Institute for Plasma Physics, EURATOM Association, Wendelsteinstr. 1, 17491 Greifswald, Germany}
\affiliation{Max-Planck/Princeton Research Center for Plasma Physics}
\author{P. Xanthopoulos}
\affiliation{Max Planck Institute for Plasma Physics, EURATOM Association, Wendelsteinstr. 1, 17491 Greifswald, Germany}
\affiliation{Max-Planck/Princeton Research Center for Plasma Physics}
\author{J. W. Connor}
\affiliation{Euratom/CCFE Fusion Association, Culham Science Centre, Abingdon, Oxon, UK, OX14 3DB}

\begin{abstract}
We investigate the linear theory of the ion-temperature-gradient (ITG) mode, with the goal of developing a general understanding that may be applied to stellarators.  We highlight the Wendelstein 7X (W7-X) device.  Simple fluid and kinetic models that follow closely from existing literature are reviewed and two new first-principle models are presented and compared with results from direct numerical simulation.  One model investigates the effect of regions of strong localized shear, which are generic to stellarator equilibria.  These ``shear spikes'' are found to have a potentially significant stabilizing affect on the mode; however, the effect is strongest at short wavelengths perpendicular to the magnetic field, and it is found to be significant only for the fastest growing modes in W7-X.  A second model investigates the long-wavelength limit for the case of negligible global magnetic shear.  The analytic calculation reveals that the effect of the curvature drive enters at second order in the drift frequency, confirming conventional wisdom that the ITG mode is slab-like at long wavelengths.  Using flux tube simulations of a zero-shear W7-X configuration, we observe a close relationship to an axisymmetric configuration at a similar parameter point.  It is concluded that scale lengths of the equilibrium gradients constitute a good parameter space to characterize the ITG mode.  Thus, to optimize the magnetic geometry for ITG mode stability, it may be fruitful to focus on local parameters, such as the magnitude of bad curvature, connection length, and local shear at locations of bad curvature (where the ITG mode amplitude peaks).
\end{abstract}

\maketitle

\section{Introduction}

The ion-temperature gradient (ITG) mode remains the chief candidate for driving turbulent transport of heat in tokamaks and stellarators.  Although the mode has been exhaustively studied in the tokamak context, the problem is still relatively fresh in the stellarator context.  Furthermore, stellarators present a unique opportunity for optimizing turbulent transport via novel magnetic equilibria \cite{mynick-stelopt}, which calls for a basic understanding of the physics that governs the ITG mode.  Indeed the neoclassical optimization of Wendelstein 7X (W7-X) has recently been shown to fortuitously have a stabilizing effect on trapped particle modes \cite{proll-prl, resilience-pop-I, resilience-pop-II}.  Can the freedom in stellarator design be used to bring about a favorable affect on the ITG mode as well?

There has been a long history of work on the theory of ITG turbulence, and the basic physical mechanisms that control the ITG mode have already been identified and are largely common to the stellarator and the tokamak.\cite{helander-a-comparison}  Thus, it is timely to refine existing theory and compare closely with numerical simulation in order to identify those mechanisms that actually matter.  This is the goal of the present work.

We simplify our analysis at the outset by assuming adiabatic electrons and ignoring trapped particles, focusing on the classical ITG mode that is driven only by the electrostatic dynamics of passing ions.  This is an accurate approximation for a large-aspect ratio tokamak and should also apply to optimized stellarators like W7-X, for which trapped particle populations reside away from the locations of bad curvature around which turbulent intensity peaks.\cite{helander-a-comparison}

The large space of possible magnetic configurations available in stellarators seems to hold the possibility of a wide variety of potential effects controlling the ITG mode.  However, our work ultimately supports a simple and largely conventional picture of this mode, namely that local scale lengths are a sufficient parameter space to characterize the ITG mode.  Nevertheless, within this picture lies the possibility to weaken the ITG instability by shortening the effective connection lengths and optimally distributing local magnetic shear.

\section{Basic analysis}

In the present work, the analytic derivations assume that the contribution from trapped particles is negligible.  This is justified when the mode is localized to regions of weak variation in the magnitude of the mean magnetic field (\ie locations of small trapped particle fraction).  We will also use simple models of the magnetic geometry.  This will simplify the analysis considerably.  The goal is to determine the basic physics controlling the ITG mode using models that are directly derived from the gyrokinetic equation with clearly stated assumptions.
 
Our starting point is the linear electrostatic gyrokinetic equation in the familiar eikonal form.  We define $g_i$ to be the non-adiabatic part of $\delta f_i$ and $\phi$ to be the electrostatic potential.  Then for the linear calculation we use the representation\footnote{Note that the eikonal representation is only appropriate when variation of the mode occurs on a smaller scale than that of equilibrium quantities like magnetic curvature and shear.  Although it is unknown to what extent this is satisfied in the radial direction, recent full-flux-surface simulations have revealed a loss of scale separation in the binormal direction for modes with small $k_\perp$.  However, details of this will be left for a future paper.} $g_i = \hat{g}_i({\bf \ell}) \exp(i S - i \omega t)$ and $\phi = \hat{\phi}({\bf \ell}) \exp(i S - i \omega t)$ to form a ``twisted slicing'' mode.\cite{roberts-taylor}  Fast oscillatory behavior is contained in the factor $\exp (i S)$, while the required anisotropy condition ${\bf B} \cdot \bnabla S = 0$ is satisfied by taking $\partial S/\partial \ell = 0$.  It is convenient to then use flux coordinates to express the perpendicular wavenumber as

\begin{equation}
\bnabla S \equiv {\bf k}_{\perp} = k_{\alpha}\bnabla\alpha + k_{\psi}\bnabla\psi,\label{kperp-eqn}
\end{equation}

\noindent where ${\bf B} = \bnabla \psi \times \bnabla \alpha$ and the wavenumbers $k_{\alpha}$ and $k_{\psi}$ are simply constants.  Thus the variation of ${\bf k}_{\perp}$ in $\ell$ is due entirely to the behavior of the known geometric functions $\bnabla\alpha$ and $\bnabla\psi$.  This sets the stage for the ballooning theory of micro-instabilities.\cite{cht-iii}  We henceforth drop the hats, substituting $\hat{g}_i \rightarrow g$ and $\hat{\phi} \rightarrow \phi$.  The linear gyrokinetic equation for the ions is

\begin{equation}
i\vpar \frac{\partial g}{\partial \ell} + (\omega - \odt)g = \varphi J_0(\omega - \ost)f_0\label{gk-eqn}
\end{equation}

\noindent where we have adopted the following definitions: $J_0 = J_0(k_{\perp}v_{\perp}/\Omega) = J_0(k_{\perp}\rho\sqrt{2}v_{\perp}/\vth)$; the thermal velocity is $\vth = \sqrt{2T/m}$ and the thermal ion Larmor radius is $\rho = \vth/(\Omega\sqrt{2})$; $n$ and $T$ are the background ion density and temperature; $q$ is the ion charge; $\varphi = q\phi/T$ is the normalized electrostatic potential.  Assuming Boltzmann electrons, the quasineutrality condition is

\begin{equation}
\int d^3{\bf v} J_0 g = n(1 + \tau) \varphi,\label{qn-eqn}
\end{equation}

\noindent where $\tau = T_i/(ZT_e)$ with the charge ratio defined as $Z = q_i/q_e$.  The equilibrium distribution is the Maxwellian

\begin{equation}
f_0 = \frac{n}{(\vth^2\pi)^{3/2}}\exp(-v^2/\vth^2),
\end{equation}

\noindent and we introduce the velocity-dependent diamagnetic frequency

\begin{equation}
\ost = \os[1 + \eta(v^2/\vth^2 - 3/2)]
\end{equation}

\noindent where $\eta = d\ln T/d\ln n$ and $\os = (Tk_{\alpha}/q)d\ln n/d\psi$.  We will also use the notation $\osT = \eta \os = (Tk_{\alpha}/q)d\ln T/d\psi$.  The magnetic drift frequency is $\odt = {\bf v}_d\cdot{\bf k}_{\perp}$ and the magnetic drift velocity is ${\bf v}_d = \hat{\bf b}\times((v_{\perp}^2/2)\bnabla \ln B  + \vpar^2\bkappa)/\Omega$, where $\hat{\bf b} = {\bf B}/B$ and $\bkappa = \hat{\bf b}\cdot\bnabla\hat{\bf b}$.  For simplicity, we shall take $\bnabla \ln B = \bkappa$ (small $\beta$ approximation).  We may then define a velocity-independent drift frequency via

\begin{equation}
\odt = \od\left[\frac{\vpar^2}{\vth^2} + \frac{\vperp^2}{2\vth^2}\right],
\end{equation}

\noindent where the drift frequency generally varies along the field line $\od = \od(\ell)$.

\subsection{Discussion of magnetic shear}\label{shear-disc-sec}

Following convention (see \ie \citet{dewar-glasser}) we can use toroidal coordinates in which the field lines are straight, writing $\alpha = \theta - \iota \zeta$, where $\iota$ is the rotational transform and $\zeta$ and $\theta$ are toroidal and poloidal angles.  We then have 

\begin{equation}
{\bf k}_{\perp} = k_{\alpha}(\bnabla\theta - \iota\bnabla\zeta - \iota'\zeta\bnabla\psi) + k_{\psi} \bnabla\psi.\label{kperp-eqn2}
\end{equation}

\noindent The affect of global shear can be identified with the term proportional to $\iota' = d\iota/d\psi$, the only term that is secular (and thus not periodic in $\zeta$).  The other terms proportional to $k_{\alpha}$ may be considered to contain ``local shear'' effects.  However, these distinctions are not necessarily meaningful to the ITG mode itself, which simply responds to the overall variation of ${\bf k}_{\perp}$ in $\ell$.  Magnetic shear thus enters our problem in two places in Eqn.~\ref{gk-eqn}, namely in the argument of $J_0$ and in the drift frequency $\od$.  

Generally speaking, as one follows a field line, magnetic shear causes field lines in a neighboring surface to drift further apart, leading to growth in the gradient of the field line label, $\bnabla\alpha$.  For finite global shear, this implies that even if $k_{\perp}^2\rho^2$ is small at a particular location along the field line, sufficiently far away the magnitude $k_{\perp}$ will become large and cause ``FLR suppression'' of the ITG mode via the $J_0$ in Eqns.~\ref{gk-eqn} and \ref{qn-eqn}.  This effect is independent of the sign of the magnetic shear.  On the other hand, where $k_{\perp}^2\rho^2$ is small ($J_0 = 1$ in Eqns.~\ref{gk-eqn} and \ref{qn-eqn}) the twisting of the mode due to magnetic shear (\ie the variation of ${\bf k}_{\perp}$ due to shear) affects the mode in the same way that variation in the curvature $\bkappa$ does since both cause variation in the function $\od(\ell)$.  As a consequence, a mode can be localized equivalently by curvature, global shear, or local shear.

To illustrate this, consider the model ``$s$-$\alpha$'' tokamak equilibrium at low $\alpha$:  the drift velocity and wavenumber can be written in terms of the poloidal angle $\theta = \pi\ell/\Lpar$ ($\Lpar$ the connection length) as ${\bf v}_d = v_{d0}(\hat{\bf y} \cos(\theta) + \hat{\bf r} \sin(\theta))$ and ${\bf k}_{\perp} = k_y(\hat{\bf y} + \hat{\bf r} \hat{s} \theta)$.  Expanding the expression $\odt = {\bf k}_{\perp}\cdot{\bf v}_d$ for small $\theta$, we obtain $\od = \odo [1 - (1/2-\hat{s}) \theta^2]$, from which we can identify a ``twist parameter'' of $1/2 - \hat{s}$, as found by \citet{kim-wakatani}.  This result expresses the very general fact that in the region surrounding a local curvature extremum, the effect of (moderate) shear is to simply broaden or narrow the confining drift-frequency well (henceforth to be called the ``drift well'') for positive and negative shear respectively.  More generally, the magnetic drift velocity maintains a roughly vertical direction when curvature is in the direction of the major radius, and so it tilts away from the flux surface as one follows a field line away from the outboard midplane.  Positive shear causes the mode to simultaneously tilt, maintaining alignment with the drift velocity and broadening the drift well associated with $\od$.  This physical effect was first recognized by \citet{choi-horton} and further elucidated by \citet{antonsen}.

\begin{figure}
\includegraphics[width=0.95\columnwidth]{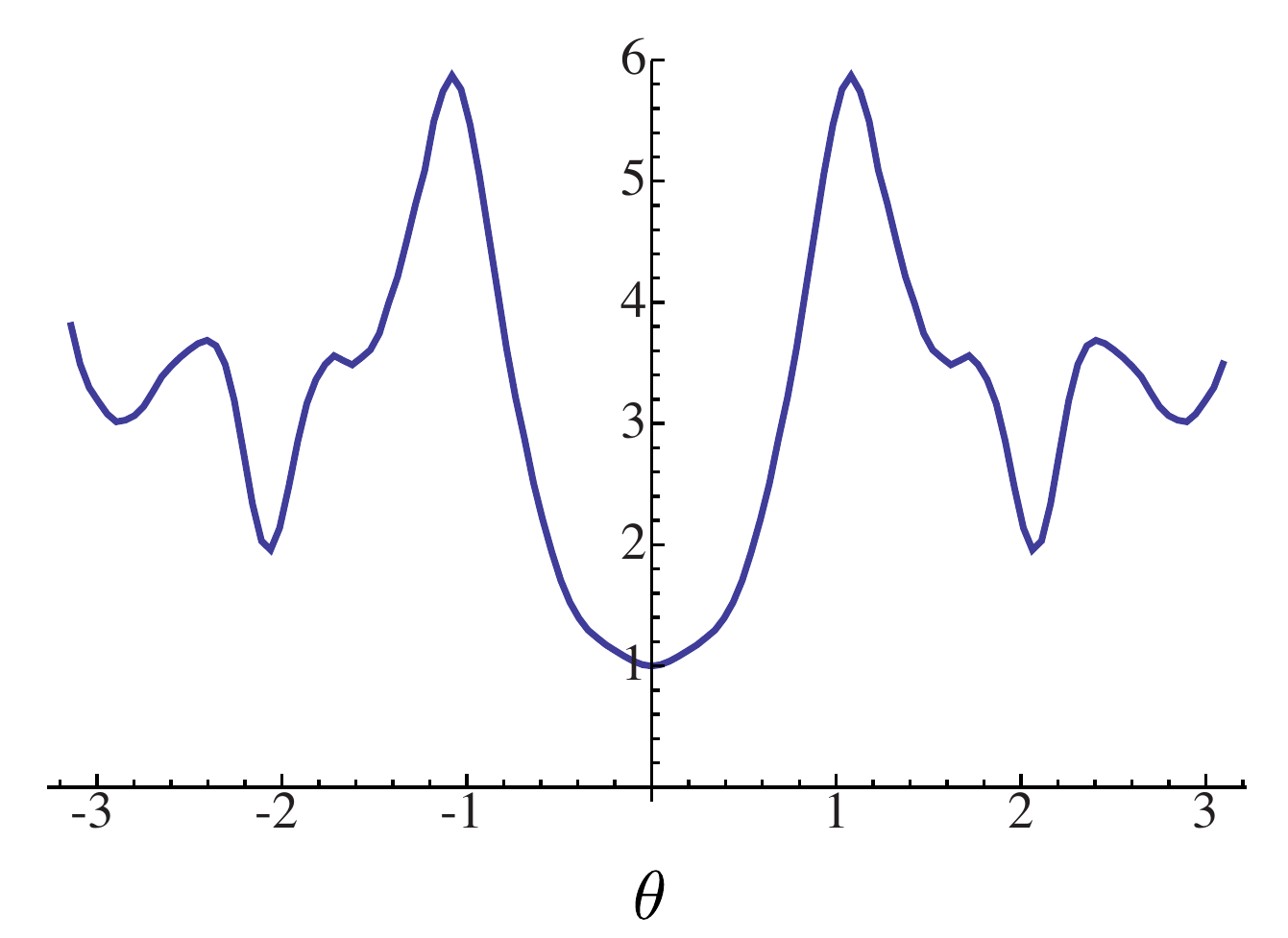}
\caption{Local shear and amplification of $|k_{\perp}|$.  The amplification factor is defined as $A(\theta) = |k_{\perp}(\theta)|/|k_{\perp}(0)| = |\bnabla\alpha|/|\bnabla\alpha|_{\theta = 0}$.  The magnetic coordinates are such that $\bnabla\alpha$ is perpendicular to the radial direction at $\theta = 0$, and ${\bf k}_{\perp}$ chosen in the surface at $\theta = 0$ (\ie $k_{\psi} = 0$).  Note that this flux tube is the same as that used in Fig.~\ref{eigenmode-localize-compare-fig}.}
\label{local-shear-vs-amp-fig}
\end{figure}

Local shear can also cause a large jump in $\bnabla\alpha$.  While the action of global shear on ${\bf k}_{\perp}$ is gradual, a sudden increase in local shear can ``box-in'' the mode by introducing hard boundaries.  As an example of this, consider the plot in Fig.~\ref{local-shear-vs-amp-fig}, in which the calculation of the amplification of $|k_{\perp}|$ is done explicitly along a W7-X flux tube.  We investigate the consequences of these ``shear spikes'' in detail in Sec.~\ref{shear-well-sec}.

\section{Local dispersion relation}\label{local-disp-sec}

The local dispersion relation is obtained when the background does not depend on $\ell$ and so a single Fourier mode can be taken in the parallel direction, \ie $g(\ell) \sim \exp(i \kpar \ell)$.  In this case the perpendicular wavenumber ${\bf k}_{\perp}$ is constant, and the diamagnetic frequency can be rewritten $\os = k_y \rho\vth/(L_n\sqrt{2})$, where $L_n = (d\ln n/ d r)^{-1}$ is the density gradient scale length and $k_y = k_{\alpha}B_0(dr/d\psi)/\sqrt{2}$ is the poloidal wavenumber.  Likewise, the magnetic drift frequency is a constant $\od = \odo = \sqrt{2}(k_y \rho)\vth/\Reff$, where we have expressed it in terms of an effective radius of curvature $\Reff$.  (This expression for the drift frequency corresponds to the ``outboard mid-plane'' scenario where the drift velocity is parallel to the diamagnetic velocity at the location of the ITG mode.)  Combining Eqns.~\ref{gk-eqn} and \ref{qn-eqn} we obtain

\begin{equation}
0 = 1 + \tau - \frac{2}{\sqrt{\pi}}\int_0^{\infty} \xperp d \xperp\int_{-\infty}^{\infty} d\xpar \left[\frac{\omega - \ost}{\omega - \odt - \xpar \op} \right]J_0^2(\xperp\sqrt{2b})\exp(-\xpar^2-\xperp^2),\label{full-disp-eqn}
\end{equation}

\noindent where $b = k_{\perp}^2\rho^2$ and $\op = \kpar\vth$, and we have used normalized velocity variables $\xperp = \vperp/\vth$ and $\xpar = \vpar/\vth$.  Taking $\kpar$ to correspond to the characteristic variation of the background (\ie the typical connection length estimate $\kpar \approx 1/qR$) is a common practice, but formally incorrect.  However, it does at least capture qualitatively an effect associated with imposing parallel variation on the ITG mode, \ie suppression via Landau damping.  If the transit frequency is sufficiently small the local dispersion relation is correct for the toroidal ITG mode (with $\kpar = 0$), as we will explicitly show in Sec.~\ref{small-wt-sec}.

\subsection{Slab ITG mode}\label{slab-sec}

The slab branch ($\od = 0$, $\op \neq 0$) is in some sense an ITG mode in free space, since it is not bound along the field line to a particular local source of free energy, and the wavenumber $\kpar$ is a free parameter.  The dispersion relation may be evaluated in terms of the plasma dispersion function and the special functions $\Gamma_0 = \Gamma_0(b)$ and $\Gamma_1 = \Gamma_1(b)$ (see Appx.~\ref{weber-int-appx}).  We find

\begin{equation}
0 = 1 + \tau + \Gamma_0\left( \xi Z(\xi) + \frac{\os}{\op}\left[\left(\frac{3\eta}{2} - 1\right)Z(\xi) - \xi \eta(1 + \xi Z(\xi))\right]\right) - \frac{\os}{\op} \eta ((1-b) \Gamma_0 + b \Gamma_1) Z(\xi),\label{slab-disp-eqn}
\end{equation}

\noindent where $\xi = \omega/\op$.  By taking the limit $\Im[\xi] \rightarrow 0+$ an instability criterion may be derived (see \citet{kadomtsev-pogutse} and Appx.~\ref{slab-mode-appx}), which for positive $\eta$ is

\begin{equation}
\eta > \frac{1}{1 + 2b(1 - \Gamma_1/\Gamma_0)}\left[ 1 + \sqrt{1+ \frac{2(1 + \tau)(1+ \tau - \Gamma_0)}{\Gamma_0^2\os^2/\op^2}\left(1+2b(1-\Gamma_1/\Gamma_0)\right)}\right]\label{slab-instability-criterion-eqn}
\end{equation}

\begin{figure}
\includegraphics[width=0.95\columnwidth]{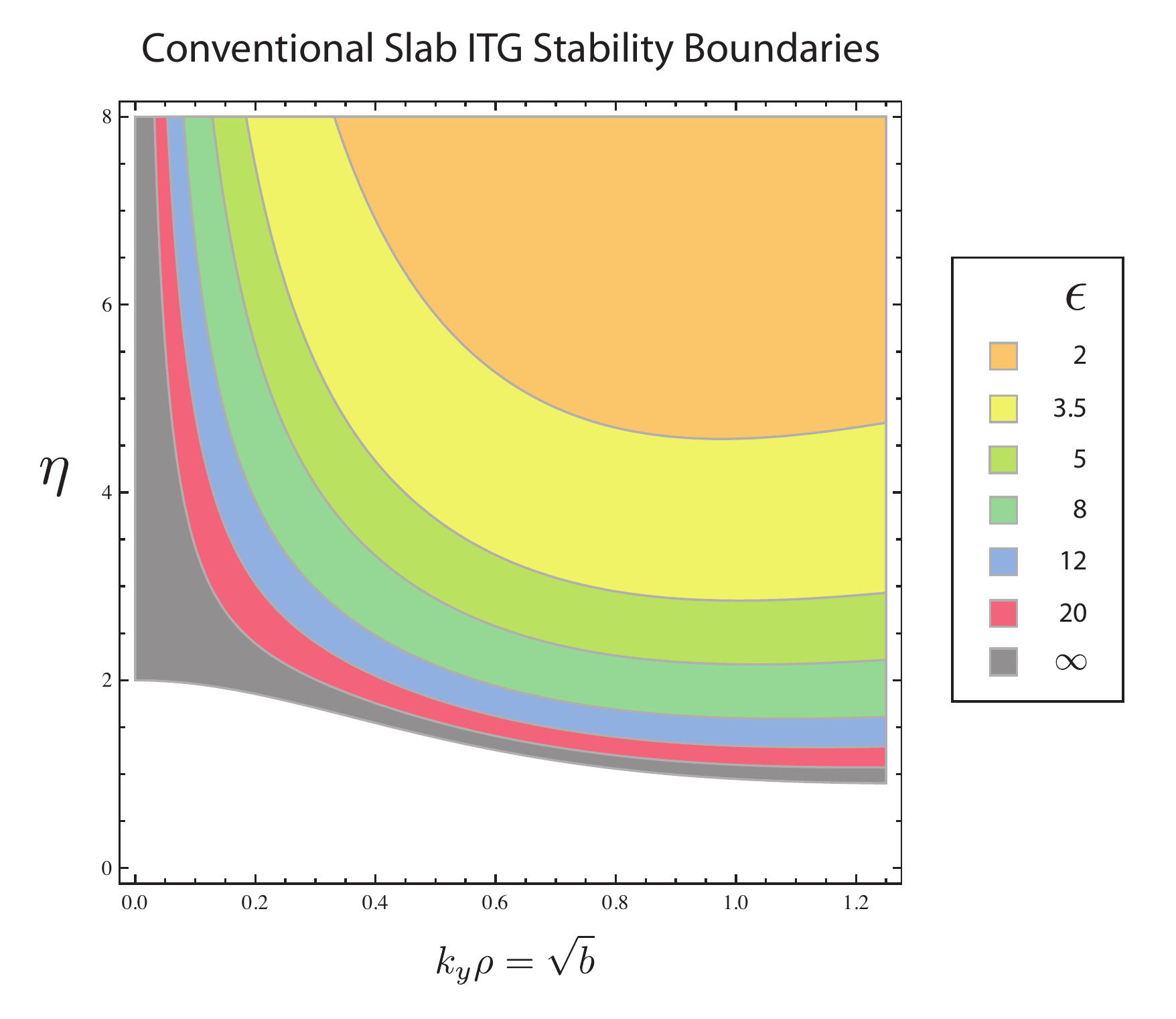}
\caption{Slab mode marginal stability contours determined by Eqn.~\ref{slab-instability-criterion-eqn}.  Note that $\epsilon = \pi/(\kpar L_n)$}
\label{eta-b-stability-fig}
\end{figure}

\noindent Fig.~\ref{eta-b-stability-fig} shows the stability diagram for several values of the parameter $\epsilon = \pi/(\kpar L_n)$.  Taking $b \rightarrow 0$ while keeping $\os$ finite, we can solve for the long-wavelength instability criterion.  Denoting the critical wavenumber as $\kyc$ and recalling that $\os = k_y\rho\vth/(\sqrt{2}L_n)$ we find

\begin{equation}
\kyc = 2\kpar L_n \sqrt{\frac{\tau(1 + \tau)}{\eta(\eta - 2)}}.\label{lw-criterion}
\end{equation}

This criterion is physically important because it sets the minimum perpendicular wavenumber needed for instability (given fixed parameters $\op$, $\tau$ and $\eta$), and therefore sets the largest scale of the ITG turbulence.  Furthermore it is a universal instability criterion insofar as the long-wavelength ITG mode is slab-like, as discussed in Sec.~\ref{gen-disp-sec} below.  Note that it implies that the minimum $k_y\rho$ is $\sim \kpar L_T$ for large $\eta$, with $L_T = (d\ln T/dr)^{-1}$.

\subsection{Toroidal ITG mode}

An analysis of the toroidal branch ($\od \gg \op$) was made by \citet{biglari} yielding a necessary condition for instability ($\eta > 2/3$).  However, the toroidal branch instability criterion is more commonly set by the parameter $\kappa = 2\osT/\odo = \Reff/L_T$ (recall the definition $\osT = \eta\os$) which measures the relative magnitudes of the temperature gradient and magnetic curvature.  The critical value $\kappa_c$ must be determined numerically, and is dependent on $\kpar$: a shorter connection length allows for a greater temperature gradient before the toroidal ITG branch sets in.

\subsection{Strongly-driven (non-resonant) limit}\label{nr-disp-sec}

Let us turn to a limit that is analytically tractable, which we will call the strongly driven limit.  We expand in $\omega/\osT = \delta \ll 1$, and order $\os/\omega \sim 1$ and $\od/\omega \sim \op^2/\omega^2 \sim k_{\perp}^2\rho^2 \sim \delta$.  In this limit, the resonant integrand in the dispersion relation may be Taylor expanded (neglecting exponentially small corrections).  The dominant term ($\sim \osT$) is zero upon integrating over velocity and what remains is a balance between zeroth order terms that form the following cubic dispersion relation

\begin{equation}
\tau \omega^3 + \left(1- b\eta\right)\os\omega^2 + \odo\osT \omega + \frac{1}{2}\osT\op^2 = 0.\label{strong-local-disp-eqn}
\end{equation}

Neglecting FLR ($b = 0$) and taking the limit $\op^2/\omega^2 \ll \odo/\omega$, one obtains the toroidal\cite{biglari} branch $\omega = (-\os \pm \sqrt{\os^2 - 4\tau\odo\osT})/(2\tau)$; the limit $\op^2/\omega^2 \gg \odo/\omega$ yields the slab\cite{cowley-kulsrud} result $\tau\omega^3 + \os\omega^2 + \osT\op^2/2 = 0.$  Note that since both $\od$ and $\os$ are linear in $k_y$, so is the solution $\omega$ for the toroidal branch.  However, the slab branch growth rate scales as $\osT^{1/3} \propto k_y^{1/3}$, which explains why the toroidal branch generally dominates over the slab branch for sufficiently large $k_y$.

\noindent Eqn.~\ref{strong-local-disp-eqn} gives a simple picture of the ITG mode but will not be quantitatively correct for realistic experimental parameters because the expansion parameter is unlikely to be very small as one can verify {\it a posteriori}.  For example, the solution for the toroidal branch $\omega \sim \sqrt{\osT\odo}$ implies that $\delta \sim \omega/\osT \sim \sqrt{1/\kappa}$, where the quantity $\kappa = \Reff/L_T \geq 4$ is the conventional toroidal ITG instability threshold parameter; \ie, $\delta$ is not so small.  Likewise for the slab branch we have $\delta \sim \op^2/\omega^2 \sim (\kpar L_T/b^{1/2})^{2/3}$ and $b \sim \delta$, so we find $\delta \sim (\kpar L_T)^{1/2}$, which is a similarly poor expansion parameter for the case $\kpar \sim 1/(q R)$.  Nevertheless, one can confirm that the numerical solutions of the general dispersion do tend asymptotically to the solutions of Eqn.~\ref{strong-local-disp-eqn}.

\subsection{General lessons from the dispersion relation}\label{gen-disp-sec}

The ordering $\kappa \gg 1$, or equivalently $\osT \gg \od$, should be applicable when the toroidal ITG mode is sufficiently unstable since $\kappa$ is the instability parameter for the toroidal branch.  It should be a more accurate approximation if the connection length is small, \ie $\Reff \kpar \gg 1$, since it is in this limit that the threshold $\kappa_c$ becomes large.  We can relate the stellarator and tokamak cases by introducing the parameter $\qeff = 1/(\Reff \kpar) = \Lpar/(\Reff\pi)$.  For W7-X this parameter can be approximated as $\qeff = 0.3$; in Sec.~\ref{saw7x-sec}, we will return to this comparison to demonstrate the similarity between the W7-X ITG mode and the ITG mode found in a circular tokamak having similar dimensionless parameters.

\begin{figure}
\includegraphics[width=0.95\columnwidth]{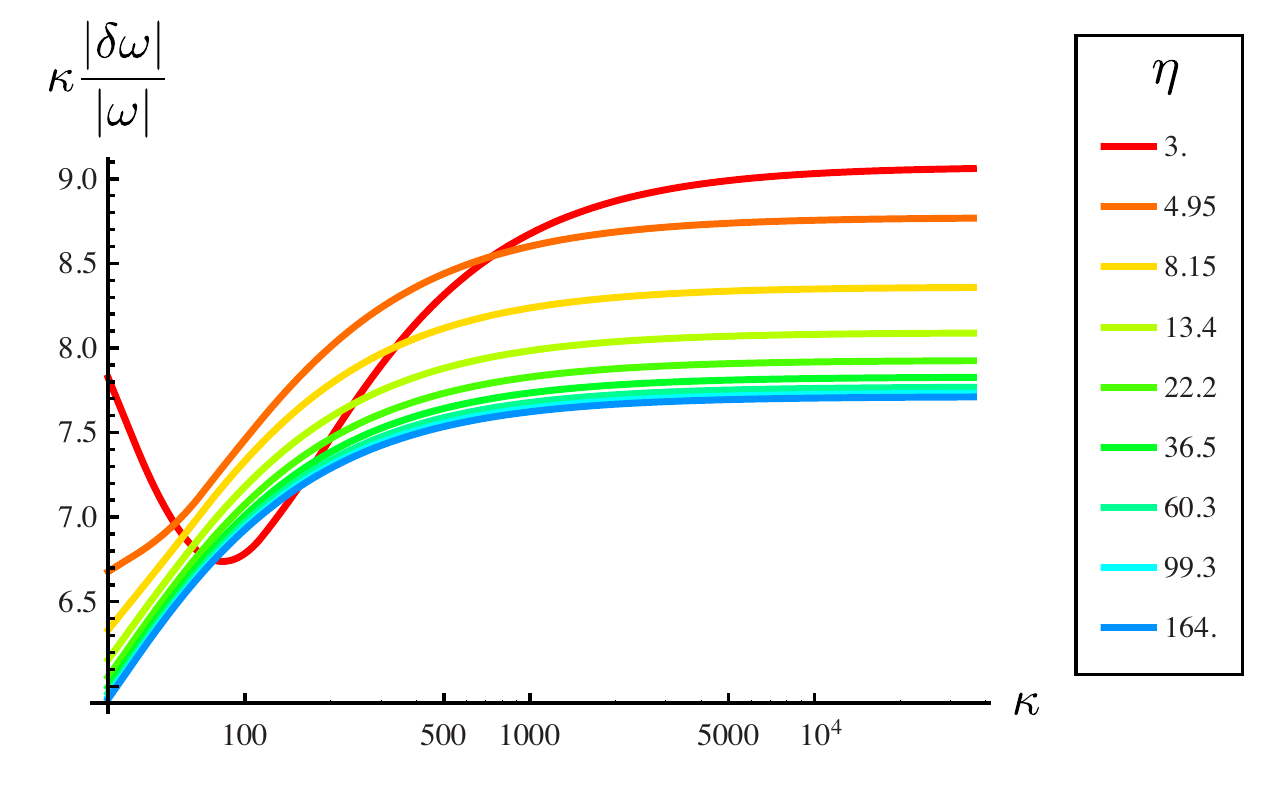}
\caption{Demonstration of slab-like behavior near long-wavelength marginal stability point: $\omega$ is the full solution of Eqn.~\ref{full-disp-eqn} and $\delta \omega$ denotes the deviation of the pure slab ($\od = 0$) solution from this solution, where the frequency is computed at twice the cutoff wavenumber $\kyc$ of the slab mode.  A range of $\eta$ and $\kappa$ are calculated to demonstrate the smallness of $\delta \omega$ and the asymptotic behavior $\delta \omega \sim 1/\kappa$ in the large-$\kappa$ limit.}
\label{long-wavelength-slaby-fig}
\end{figure}

Though moderately large $\kappa$ may be achievable with a small connection length, the condition $\op \ll \omega$, used to derive Eqn.~\ref{strong-local-disp-eqn}, cannot be satisfied for all $k_y$.  This is because both $\os$ and $\od$ scale linearly with $k_y\rho$ whereas the smallness of $\op$ is limited by the magnetic configuration.  For sufficiently small $k_y$ the frequencies must balance, $\op \sim \omega$, at which point the expansion of the parallel resonance will no longer be valid and the correction due to $\odo$ will become formally small relative to the resonant integral.  However, the dispersion relation in the limit $\omega \sim \op, \osT \gg \od$ is the slab dispersion relation (since we neglect magnetic trapping) and we can refer to the analysis of the slab mode to determine the long-wavelength stabilization criterion, \ie Eqn.~\ref{lw-criterion} will apply.  Numerical solutions of the full dispersion relation Eqn.~\ref{full-disp-eqn} confirm that near the marginally stable wavenumber $\kyc$ the slab dispersion relation is a good approximation to the full dispersion relation for modestly large $\kappa$, and the error scales asymptotically as $1/\kappa$ for large $\kappa$ (see Fig.~\ref{long-wavelength-slaby-fig}).

In summary, the large-$\kappa$/small-$\qeff$ limit yields an ITG mode that is slab-like for a range of long-wavelength modes that satisfy the ordering $\od \ll \omega, \osT, \op$, and is in this sense more ``slab-like'' than a typical tokamak scenario.  Evidence from W7-X linear simulations support this conclusion.\cite{helander-a-comparison}  We emphasize that this is not a special feature of the stellarator ITG mode, but rather is set by the value of local instability parameters ($\Reff/L_T$, $\qeff$, \etc).

\begin{figure}
\includegraphics[width=0.95\columnwidth]{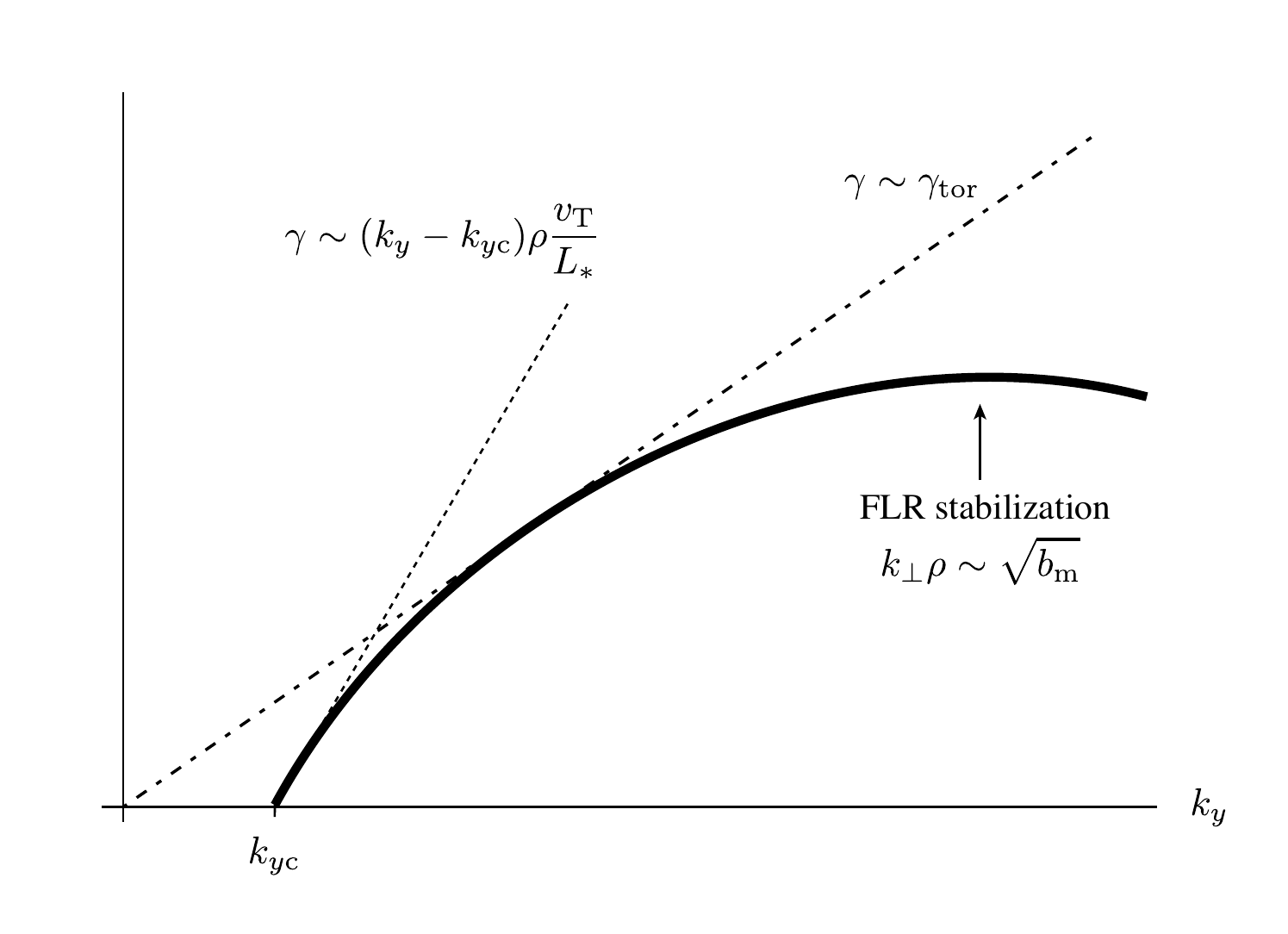}
\caption{Schematic growth rate curve of the ITG mode at $\kappa > \kappa_c$ (unstable toroidal branch) according to the local dispersion relation.  Note that $\kyc$ denotes the slab-mode cutoff wavenumber (see Eqn.~\ref{lw-criterion}), $L_*$ tends to $L_T$ at large $\eta$, and $\gamma_{\mathrm{tor}}$ is the imaginary part of the solution of Eqn.~\ref{strong-local-disp-eqn} with $\op = 0$.}
\label{local-disp-gamma-fig}
\end{figure}

We summarize the features of the toroidally destabilized ($\kappa > \kappa_c$) ITG growth rate curve in Fig.~\ref{local-disp-gamma-fig} according to the local dispersion relation.  Stabilization occurs at long wavelength due to a balance between drive ($\osT$) and Landau damping ($\op$); a weakly unstable slab-like ITG mode is found just above the cutoff wavenumber $\kyc$; a slab-like ITG mode continues at higher $k_y$ and transitions to a linear scaling $\gamma \propto k_y\rho$ that can approach the limit of strongly driven toroidal ITG mode.  At larger $k_y \rho$, the growth rate falls below a linear scaling under the influence of FLR (the term containing $b$ in Eqn.~\ref{strong-local-disp-eqn}); this can be estimated to occur at some fraction of the maximum unstable $b = \bc$, calculated by setting the discriminant of Eqn.~\ref{strong-local-disp-eqn} to zero with $\op = 0$: $\bc = \eta^{-1} + \sqrt{\tau\odo/\osT}$.  At sufficiently high $\kappa$, however, the ITG mode can resist absolute stabilization at $b \sim \bc$ and can undergo further peak(s) in growth rate due to higher-order FLR terms; this is called the short wavelength ITG\cite{switg} mode and is regarded as being relatively unimportant for transport as compared with its longer-wavelength counterpart.

\section{Nonlocal analysis}

\subsection{Small transit frequency approximation}\label{small-wt-sec}

Let us now consider the (realistic) case of a non-uniform background plasma.  Adopting the ordering scheme of the previous section, we may consider a limit in which the transit frequency is small relative to the mode frequency, but has a zeroth-order effect on the solution.  For small transit frequency, the gyrokinetic system may be reduced to a second order differential equation (see Appx.~\ref{small-wt-appx})

\begin{equation}
(1+\tau) \varphi = \mathcal{L}\varphi,\label{cht-eqn}
\end{equation}

\noindent where

\begin{equation}
\mathcal{L} = \frac{2}{\sqrt{\pi}}\int_0^{\infty} \xperp d \xperp\int_{-\infty}^{\infty} d\xpar J_0^2(\xperp\sqrt{2b})\exp(-\xpar^2-\xperp^2) \left[1 - \left(\frac{\ot\xpar}{\omega - \odt}\frac{\partial}{\partial \vartheta}\right)^2 \right] \frac{(\omega - \ost)}{(\omega - \odt)}
\end{equation}



\noindent We have introduced the coordinate $\vartheta = \pi \ell/\Lpar$ to denote an angle-like variable that takes values $-\pi$ and $\pi$ at the peaks in $\od$ around a well.  It is distinct from the poloidal coordinate $\theta$, except for simple axisymmetric cases.  Thus $\Lpar$ is the parallel connection length and the corresponding transit frequency is $\ot = \vth\pi/\Lpar$.  This equation is similar in form to what is obtained by expanding the parallel resonance in the local dispersion relation, but it is superior because it enables us to solve for the variation of the mode along the field.  Closely related equations were given by \citet{cht-iii} (Eqn.~35), \citet{antonsen-lane} (Eqns.~39a-c), and \citet{horton-choi-tang} (Eqn.~1), all employing a small-transit-frequency expansion.  

The part of $\mathcal{L}$ that is differential in $\vartheta$ is ostensibly small in the expansion parameter $\ot/\omega$, but if we adopt the strongly driven ordering introduced in Sec.~\ref{nr-disp-sec}, namely $\omega/\osT = \delta \ll 1$ and $\od/\omega \sim \ot^2/\omega^2 \sim k_{\perp}^2\rho^2 \sim \delta$, then we find that all terms in the differential equation are the same order.  Thus Eqn.~\ref{cht-eqn} becomes

\begin{equation}
\left[\tau + (1-b\eta)\frac{\os}{\omega} + \frac{\od\osT}{\omega^2} - \frac{\osT\ot^2}{2\omega^3} \frac{\partial^2}{\partial \vartheta^2}\right] \varphi = 0.\label{cht-nonres-eqn}
\end{equation}


This is an asymptotically correct way to include the effects associated with the localization of the ITG mode along the field line; higher-order derivatives with respect to $\vartheta$ do not enter at leading order in $\ot/\omega$.  The coefficients of Eqn.~\ref{cht-nonres-eqn} can have both oscillatory behavior (\eg due to drift wells) and secular behavior due to magnetic shear.  For axisymmetric geometry with zero shear, the problem becomes Floquet-like since the drift frequency is periodic along the field line, and in the simplest scenario, a simple sinusoidal dependence of the coefficients leads to a Matheiu equation.  A further expansion around a quadratic minimum leads to a Weber equation.  These various forms have been encountered before in many works on the ITG mode.\cite{choi-horton, horton-choi-tang, romanelli-1989, romanelli-chen-1991, bhattacharjee, candy-waltz-rosenbluth}.

Let us now take a simple form for the spatial dependence of the magnetic drift $\od$ and perpendicular wavenumber ${\bf k}_{\perp}$.  For the sake of generality, we take a quadratic potential

\begin{eqnarray}
\od = \odo [1 - \lambda \vartheta^2]\label{local-od-eqn}\\
b = b_0(1 + s^2 \vartheta^2)\label{local-b-eqn}
\end{eqnarray}

\noindent where $b_0 = k_y^2\rho^2$ and an effective shear parameter $s$ has been introduced (which need not be equal to the global shear).  These forms arise from expanding a general magnetic equilibrium about the extremum of normal curvature at $\vartheta = 0$, taking $k_\psi = 0$ and expanding the function $\bnabla \alpha$ about $\vartheta = 0$ in Eqn.~\ref{kperp-eqn}.  This expression captures the physical effect of a mode centered at $\vartheta = 0$ that tilts with shear as $\vartheta$ increases, as discussed in Sec.~\ref{shear-disc-sec}.

Using the Eqns.~\ref{local-od-eqn} and \ref{local-b-eqn}, the eigenmode equation \ref{cht-eqn} is now a Weber equation.  Bound solutions (those that tend to zero as $\vartheta \rightarrow \infty$) have the form $H_n(\sqrt{\sigma}\vartheta)\exp(-\sigma \vartheta^2/2)$, where $H_n$ is a Hermite polynomial and $n = 0, 1, 2, ...$; Substituting this {\it Ansatz} into Eqn.~\ref{cht-eqn} and equating the coefficients of $\vartheta^2$ we obtain $\sigma$

\begin{equation}
\sigma^2 = -\frac{2}{\ot^2}\left(\omega\odo\lambda + b_0s^2\omega^2\right),\label{sigma-eqn}
\end{equation}

\noindent where the root satisfying $\Re[\sigma] > 0$, denoted $\sigma^+$, is chosen.   The constant terms of Eqn.~\ref{cht-eqn} then yield the dispersion relation

\begin{equation}
\tau \omega^3 + \left(1- b_0\eta\right)\os\omega^2 + \odo \osT \omega + \Delta = 0.\label{weber-disp-eqn}
\end{equation}

\noindent  This is identical to the local dispersion relation given in Eqn.~\ref{strong-local-disp-eqn} except for the final term $\Delta = \osT\ot^2(n+1/2)\sigma^+$ , which depends on the eigenmode width via $\sigma^+$.  Eqn.~\ref{weber-disp-eqn} may be efficiently solved numerically by rearranging terms and squaring to obtain a quintic in $\omega$; roots that do not obey Eqn.~\ref{weber-disp-eqn} are unphysical and can be discarded.  We find that the effect of the correction $\Delta$ can be either stabilizing or destabilizing, an observation which is somewhat surprising given that we generally expect the effect of localizing a mode along the field line to be stabilizing.

We define several regimes below by determining the relative strength of the terms for different values of $k_y \rho$.

\subsubsection{high $k_y\rho$: strongly-localized modes}

For sufficiently large wavenumbers the mode becomes strongly-localized ($\Re[\sigma] \gg 1$).  This can can be inferred by inspecting Eqn.~\ref{sigma-eqn} and noting that $\omega$ increases with $k_y$.  Two cases are considered.

\paragraph{$|s| = 0$:}

If the second term on the right of Eqn.~\ref{sigma-eqn} is negligible then the first term ensures localization of the mode but the resulting correction $\Delta$ to Eqn.~\ref{weber-disp-eqn} is small and can be dropped to yield precisely the toroidal branch of the non-resonant local dispersion relation, (\ie Eqn.~\ref{strong-local-disp-eqn} with $\op = 0$).

If we retain $\Delta$ in the equation and take $\lambda = s_0 - s$, where $s_0$ is a geometry-dependent constant, we can solve Eqn.~\ref{weber-disp-eqn} numerically.  We observe that shear destabilizes the mode ($d\gamma/ds > 0$) for $s < s_0$, as also noted by \citet{kim-wakatani} for a simple tokamak case with $s_0 = 1/2$.  By further taking small $\Delta$, one may then calculate the frequency as a correction to the pure toroidal branch: \eg, for $\os/\omega = 0$ and $n = 0$ we find $\delta \omega = (1-i)\ot \sqrt{\lambda}(\osT/\odo\tau)^{1/4}/4$.  We note that the drift well is locally inverted for positive $\lambda$ (\ie $s \geq s_0$), in which case it is implausible to assume that the mode will remain localized at $\vartheta = 0$.

\paragraph{finite $|s|$}

The second term in Eqn.~\ref{sigma-eqn} dominates and for $\Im[\omega] > 0$ we may take $\sigma^+ = -i\sqrt{2b_0}|s|\omega/\ot$, resulting in a quadratic dispersion relation with the solution

\begin{equation}
\omega = \frac{1}{2\tau}\left(-\os(1-b_0\eta) \pm \sqrt{\os^2(1-b_0\eta)^2 - 4\tau\osT[\odo - i(n+1/2)\ot |s| \sqrt{2b_0}]}\right)\label{weber-strong-local-eqn}
\end{equation}

This solution is almost the same as what is obtained from the local dispersion relation except for the term within the radical that is proportional to $|s|$.  To our knowledge, this correction has not been noticed before.  Unlike the twist parameter, this depends only on the magnitude of shear, is independent of the magnetic drift frequency, and produces greater instability at high-$n$.  However, the validity of Eqn.~\ref{cht-eqn} is limited by the condition $\vth \partial/\partial \ell \ll \omega$, which limits the allowable size of $n$.  Assuming the shear drive dominates over the toroidal drive, this condition implies $n \ll \osT/(\ot s^2) \sim {\mathcal O}(\delta^{-3/2})$.  Furthermore, the size of the shear term relative to the toroidal drive term is $(1+n/2)|s|/(2\qeff)$ where we recall that $\qeff = \vth/(\Reff\ot)$.  Thus, we expect that finite-shear destabilization can in principle dominate over (or enhance) the toroidal ITG drive and the signature of this would be the appearance of higher harmonics ($n > 0$) with oscillatory structure along the field line.

For sufficiently large $k_y$ the mode is ultimately stabilized by FLR via the $b_0\eta$ term in Eqn.~\ref{weber-strong-local-eqn}, as occurs in the local dispersion relation \ref{strong-local-disp-eqn}.

\subsubsection{moderate-$k_y$: marginally localized modes}

The applicability of Eqn.~\ref{weber-disp-eqn} is limited to the cases where the mode is sufficiently localized, $Re[\sigma^+] \gtrsim 1$; otherwise, the expansion of the magnetic drift \ref{local-od-eqn} is invalid.  If the mode is delocalized, a general form of $\od(\vartheta)$ may be used and the eigenmode equation then solved numerically as in \citet{bhattacharjee}.

Assuming the mode is localized, the $\Delta$ must dominate at sufficiently low-$k_y$ due to the first term on the right of Eqn.~\ref{sigma-eqn}.  We can demonstrate this by first assuming linear scaling $\omega \propto k_y$ for the strongly-localized mode and observing that all other terms scale more strongly in $k_y$ than this term and so must become subdominant at low $k_y$.  In this case the resulting growth rate (obtained by balancing the first and last terms in Eqn.~\ref{weber-disp-eqn}) scales like $\gamma \sim (\osT\ot)^{2/5}(\odo\lambda)^{1/5}$ and we can confirm that the $\Delta$ term is dominant.

\subsubsection{low-$k_y$: delocalized modes}

Sufficiently small $k_y$ ultimately renders both the terms in Eqn.~\ref{sigma-eqn} small, which causes the eigenmode width to diverge and invalidates the local expansion about the curvature minimum (Eqn.~\ref{local-od-eqn}).  Although Eqn.~\ref{cht-nonres-eqn} remains valid as long as transit dynamics are slow, \ie $\ot \ll \omega$, the ITG mode structure now depends on details of the ``landscape'' of curvature and shear.

However, the both the magnetic drift and FLR terms become weak in this limit (since $\omega\odo/\ot^2 \ll 1$) and so the dominant effect of geometry that remains is the two-fold periodicity of the toroidal domain.  Thus we are left with a simple slab ITG mode at zeroth-order.  (Note that the slab mode cannot be derived as a higher-mode-number solution of Eqn.~\ref{weber-disp-eqn} because the solutions $H_n(\sqrt{\sigma}\vartheta) \exp(-\sigma \vartheta^2)$ are bound modes and the slab mode is not; however the slab mode arises easily from Eqn.~\ref{cht-nonres-eqn}.)  We can refer to the simple analysis of the slab mode (reviewed in Sec.~\ref{slab-sec}) once we have specified the parameter $\kpar$.  For a tokamak, one estimates the parallel connection length as the distance from the inboard to the outboard $\Lpar = \pi qR$, which matches the distance half way around the torus in $\theta$, leading us to the familiar rule of thumb $\kpar \sim 1/(qR)$.  In a stellarator, the connection length between good and bad curvature can be much smaller than the length from the outboard to the inboard (a fortunate fact) and so the slab ITG mode need not be constrained by $\Lpar$.

Note that periodic slab-like modes are not true global modes (even in flux tube geometry), unless global magnetic shear is exactly zero so that periodicity can be exactly satisfied.  Thus there is discontinuity in the linear mode analysis at the point of zero global shear for delocalized modes, extending across a flux surface; however, modes that are sufficiently localized (strongly ballooning) should not depend sensitively on global shear.  See also \citet{mcmillan-dewar} for more on the connection between scale separation and the existence of global modes.

\subsection{Localization in Numerical Simulations}

\begin{figure}
\subfigure[Eigenmode ($k_y\rho = 0.5$) with Gaussian fit.]{\includegraphics[width=0.45\columnwidth]{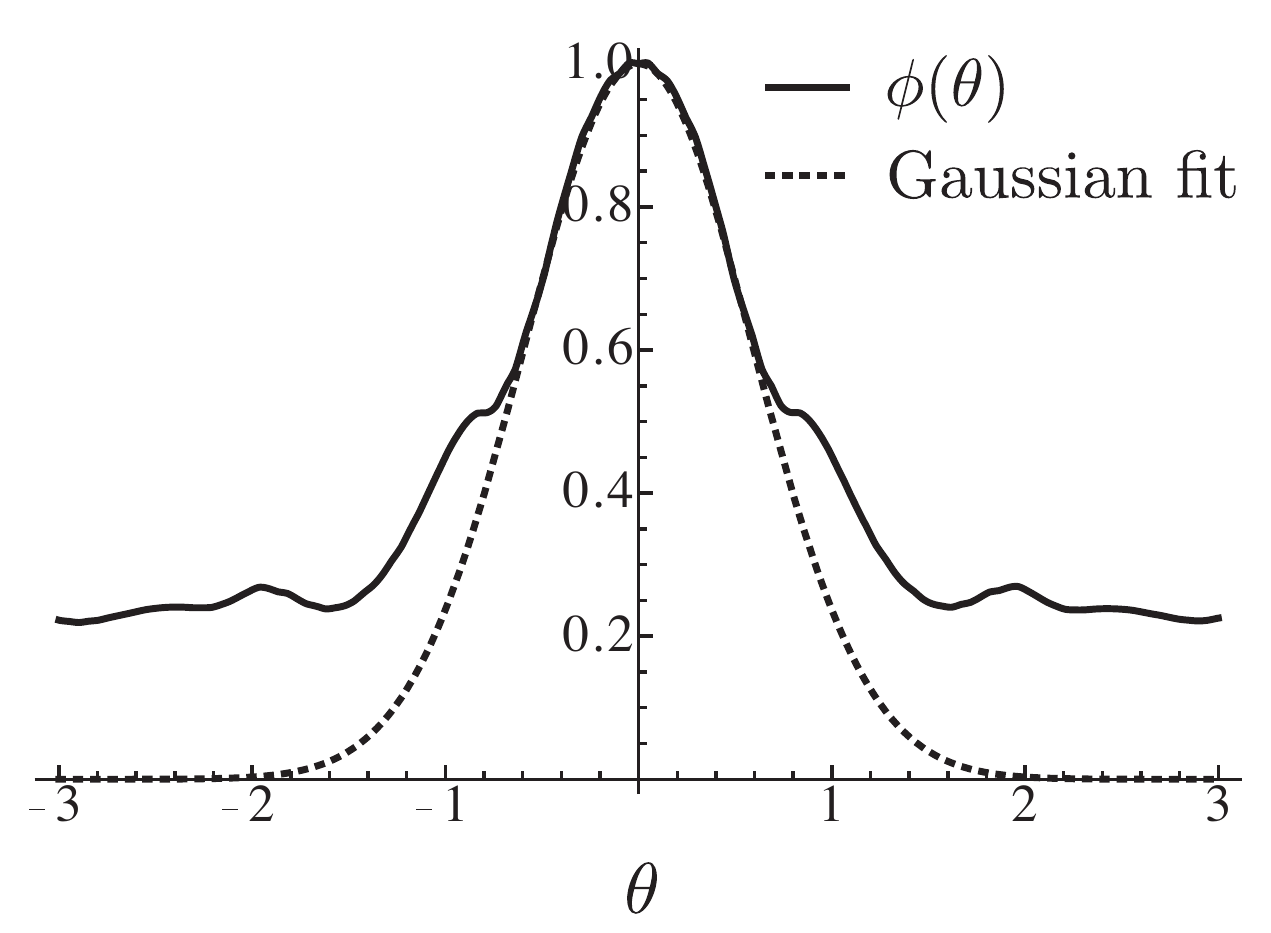}\label{eigenmode-localize-compare-fig-a}}
\subfigure[Exact drift well and {\em theoretically implied} local quadratic fit.]{\includegraphics[width=0.45\columnwidth]{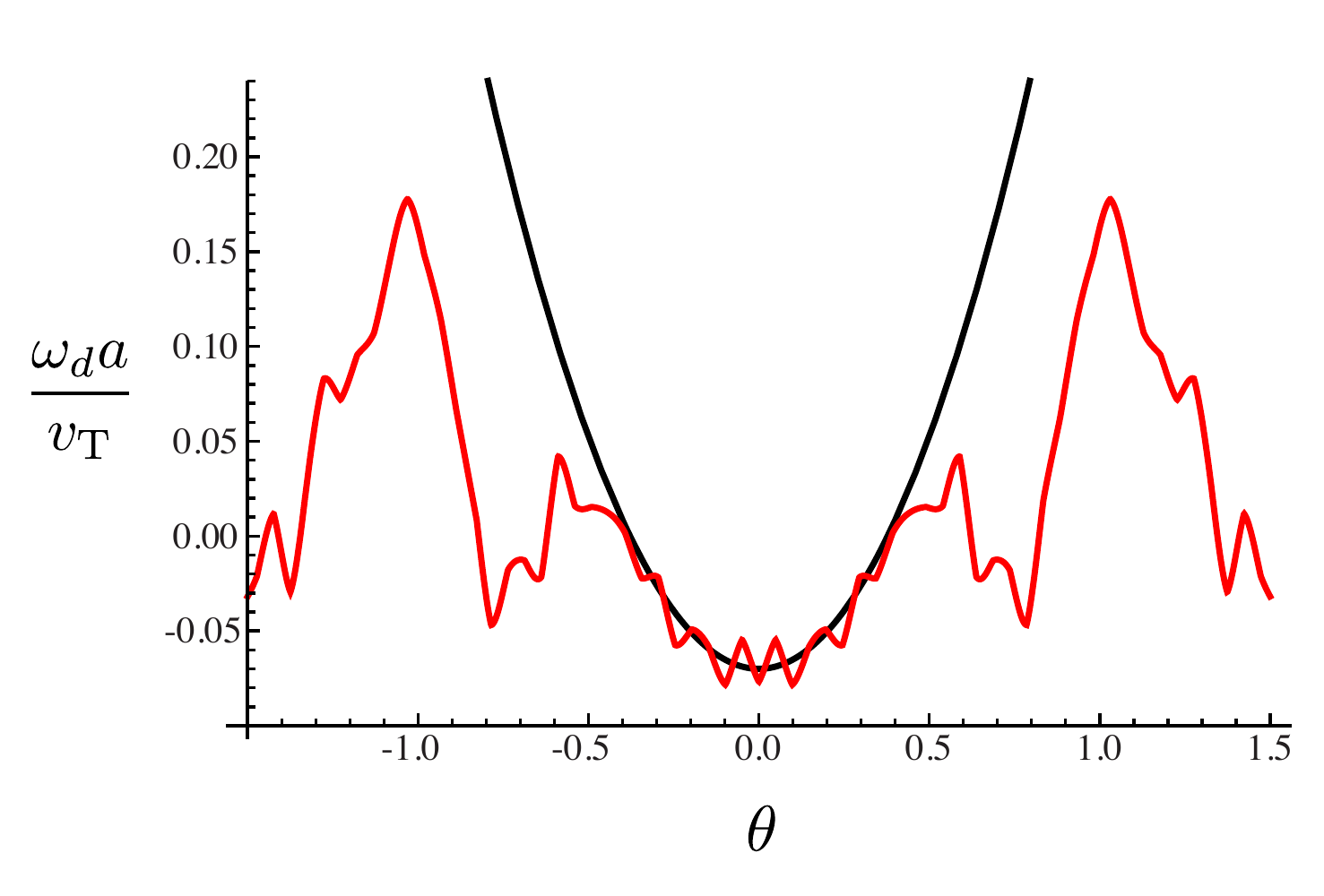}\label{eigenmode-localize-compare-fig-b}}
\caption{Comparison of theoretical and exact eigenmode localization.}\label{eigenmode-localize-compare-fig}
\end{figure}

The validity of Eqns.~\ref{sigma-eqn}-\ref{weber-disp-eqn} is limited to strongly driven (non-resonant) modes.  As shown in Sec.~\ref{gen-disp-sec}, it is not expected that the non-resonant limit will give quantitatively accurate results for realistic plasma parameters.  Nevertheless it is worthwhile to compare the theoretically predicted mode localization directly with the numerically computed exact eigenfunctions.  This comparison is shown in Fig.~\ref{eigenmode-localize-compare-fig}.  The exact eigenmodes generally exhibit some degree of peaking at the center of drift wells.  As expected, this ballooning is weak for marginally unstable modes, but becomes strong for strongly unstable modes.  A typical case for $k_y\rho = 0.5$ is show in Fig.~\ref{eigenmode-localize-compare-fig-a}; relevant parameters are $a/L_T = 2$, $a/L_n = 0$ and $a/\Lpar \approx 0.1$, where $a$ is a reference scale length (minor radius).  (More strongly ballooning cases are available, but this eigenfunction is chosen because the parameters seem experimentally accessible.)  We can compare the localization of this mode to theoretical expectation by fitting a gaussian envelope.  The width of this envelope determines $\sigma$.  Then using Eqns.~\ref{full-disp-eqn} and \ref{sigma-eqn} (\ie $\Delta \approx 0$, and $|s| \approx 0$), and taking the minimum value of the exact drift frequency (as plotted in Fig.~\ref{eigenmode-localize-compare-fig-b}) to determine $\odo$, we can determine the implied local quadratic drift well of Eqn.~\ref{local-od-eqn} (\ie, $\lambda$ can be determined).  The result is a reasonable fit to the exact drift frequency, as shown in Fig.~\ref{eigenmode-localize-compare-fig-b}

\section{Gyrokinetic first-principle models of the ITG mode}

Let us now analyze the full gyrokinetic equation using simple model geometry.  We must consider at the outset whether to look for solutions that are localized or those that extend along the field line, as the method we employ may exclude one type of mode in favor of the other.  We consider two scenarios, both concerning the limit where the parallel ion dynamics are resonant and strongly-ballooning fluid-like ITG modes are not present:

In Sec.~\ref{slab-bloch-mode-sec} we consider the long-wavelength limit of unbound (non-decaying along the field line) ITG modes in conditions of zero (or near-zero) global shear.   We call these ``Bloch'' ITG modes because of their periodic structure.  The solution is found as a perturbation to the conventional slab ITG mode, capturing the effect of a weak magnetic drift.  It is shown that this induces a ripple structure onto the slab mode, an effect that should be more visually pronounced in stellarator geometries, where the connection length between the outboard and inboard midplanes can be larger than the connection length between areas of good and bad curvature.  We also find that the correction to the frequency enters at second order in $\od/\omega$, making it weaker than the estimate obtained from expansion of the local dispersion relation.

In Sec.~\ref{shear-well-sec} we find modes in a ``square-well'' domain, bounded by spikes of infinite shear, which we call ``boxed'' ITG modes.  For zero magnetic drift these modes exhibit a very similar marginal stability criterion to a slab mode (with an appropriately chosen $\kpar$), but that they can nevertheless have significantly lower growth rates when destabilized.  This can be physically attributed to the {\em irreversible} loss of energy at the boundaries carried by outgoing particles; the reduction in the growth is observed to be approximately half the transit frequency.

\subsection{Model I: ``Nearly-slab'' Bloch wave}\label{slab-bloch-mode-sec}

As argued in Secs.~\ref{gen-disp-sec} and \ref{small-wt-sec} the ITG mode at long perpendicular wavelengths and strong drive (large $\osT/\od$) should be slab-like due to the fact the magnetic drift frequency is small compared to the other relevant frequencies, $\od \ll \omega, \ot, \osT$.  Thus we may construct our solution as a slab mode that obeys physical periodicity conditions, with a small correction due to the magnetic drift.

Let us make a few comments on the issue of magnetic shear.  FLR-induced shear is negligible ($J_0 \approx 1$) in this limit; this has been numerically confirmed for long-wavelength modes in W7-X under the influence of shear spikes (see Sec.~\ref{shear-well-sec}) and also follows from the long-wavelength behavior of Eqn.~\ref{weber-disp-eqn}.  Thus the effect of shear is confined to the magnetic drift and so will affect the mode only to ``a small degree.''  Note however that any amount of non-zero shear invalidates the periodic modes by introducing secularity that cannot necessarily be treated by perturbation theory since an arbitrarily small $\od$ will become large at sufficiently large ballooning angle.  However, \citet{romanelli-chen-1991} successfully found a weakly decaying slab mode solution using a multi-scale approach in ballooning space; see also the multi-scale analysis of \citet{bhattacharjee} and \citet{connor-hastie-2004}.  Here, we treat the global shear itself as sufficiently small such that it does not affect the frequency correction that we calculate.  We have in mind configurations with weak global shear such as W7-AS and W7-X.  In this case we need only solve the gyrokinetic equation assuming periodic modes, as discussed in Appx.~\ref{weak-shear-disc-appx}.

We should now use the geometric angle $\theta$ as the field-line following coordinate since the mode must respect the periodicity of the physical domain, and thus the mode variation is no longer determined only by the drift wells.  We would like a form of $\od$ that is simple but captures the essential physics.  For axisymmetric geometry, the curvature landscape is exactly periodic as one follows a field line (\ie in $\ell$).  However, in the non-axisymmetric case the pattern of wells never perfectly repeats unless the rotational transform is rational.  Nevertheless, the normal curvature does exhibit oscillatory behavior along a field line, and so a periodic array should be a useful paradigm to understand modes that extend across areas of both ``good'' and ``bad'' curvature.  Consider for instance the oscillatory behavior in the normal curvature in W7-X in Fig.~\ref{shear-spike-fig} (for a field line that passes through the outboard midplane at the bean-shaped cross section).  One might attribute the quasi-periodic appearance of the normal curvature to the fact that the field line cuts at an angle across a two-dimensional periodic lattice.

We thus consider a simple oscillatory form of the magnetic drift, $\od = \odo\cos(m_d\theta)$.  To model W7-X one might take $m_d = 3$.  We perform a perturbative expansion in $\od/\omega$ to find the mode frequency to second order: $\omega = \omega_0 + \omega_1 + \omega_2$.  We need not be concerned about secular evolution of the solution because strict periodicity in $\theta$ is required, and it is enforced at each order in the expansion.  The details of the derivation are in Appx.~\ref{nearly-slab-appx}.  At zeroth order, we have a simple slab mode (with parallel wavenumber $\kpar = M\ot/\vth$), as described in Sec.~\ref{slab-sec}; at first order we find that $\omega_1 = 0$; at second order we obtain a rather lengthy expression for the frequency correction, which we abbreviate as

\begin{equation}
\omega_2 = \odo^2{\mathcal I},\label{o2-soln}
\end{equation}

\noindent where ${\mathcal I}$ is given in Eqn.~\ref{o2-int}.  Thus, we have found that the drift frequency contributes at second order to the ITG mode, in contrast to the small-$\od$ expansion of the local dispersion relation where the contribution is first-order.  This is presumably due to the fact that the delocalized ITG mode extends across both good and bad curvature.

\subsubsection{Numerical experiment: modeling the W7-X ITG mode using an axisymmetric configuration}\label{saw7x-sec}

The key to modeling W7-X ITG mode with an axisymmetric configuration is setting the connection lengths to be equal.  We set the wavelength equal to $2\pi/\kpar = 2\Lpar = 2 \pi \qeff\Reff$ and find

\begin{equation}
q_{\mathrm{eff}} = \frac{\Lpar}{\pi \Reff}.
\end{equation}

\begin{figure}
\includegraphics[width=0.95\columnwidth]{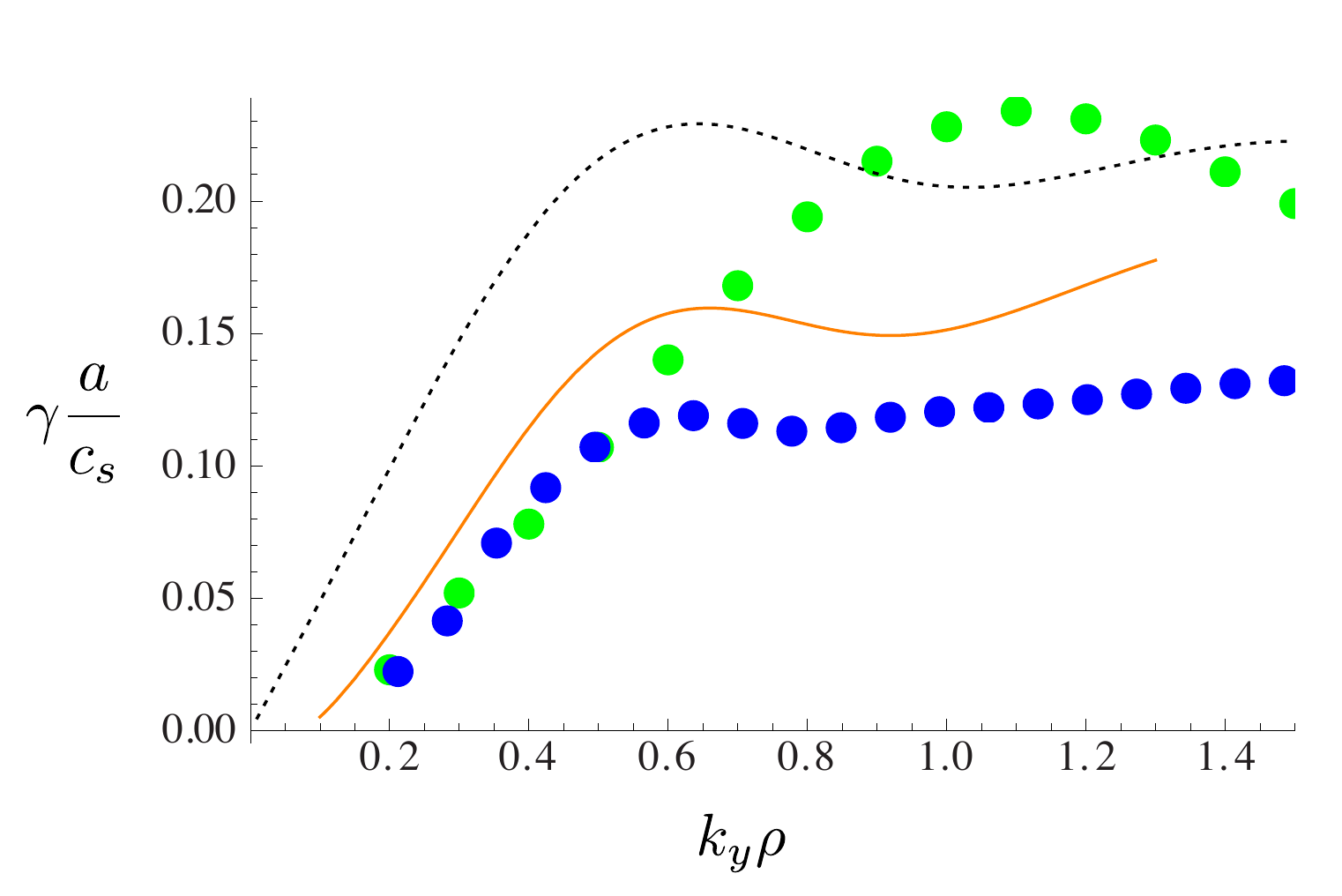}
\caption{Growth rate comparison: Green dots indicate W7-X (GENE: $a/L_T = 3$, $a/L_n = 1$, zero shear) and blue dots indicate a model tokamak geometry (GS2: $R/L_T = 31.6$, $L_n/L_T = 3$).  The dashed black line indicates a pure curvature mode ($\kpar = 0$, $R/L_T = 31.6$) and the orange line indicates a ``boxed mode'' (see Sec.~\ref{shear-well-sec}).}
\label{SAW7-X-vs-W7-X-fig}
\end{figure}

For W7-X, we take $R = 5.5 m$ and calculate $\Lpar/R = 0.96$ by integrating the numerically calculated equilibrium along the field line, $2\Lpar = \int d\ell$ between the two spikes of shear depicted in Fig.~\ref{shear-spike-fig}; this leads to $q_{\mathrm{eff}} \approx 0.3$.  These parameters are used with a model ``$s$-$\alpha$'' equilibrium with low $\alpha$ and zero global shear ($\hat{s} = 0$).  Linear simulations are performed with the gyrokinetic code GS2.  One example is given in Fig.~\ref{SAW7-X-vs-W7-X-fig}.  We note that the growth rates seem to match quite well for low-$k_y$ but diverge at high $k_y$; however, because the intensity of ITG turbulence peaks at low-$k_y$, it is significant to find such an agreement between two such different magnetic configurations.

\subsection{Model II: Boxed ITG mode}\label{shear-well-sec}

\begin{figure}
\includegraphics[width=0.95\columnwidth]{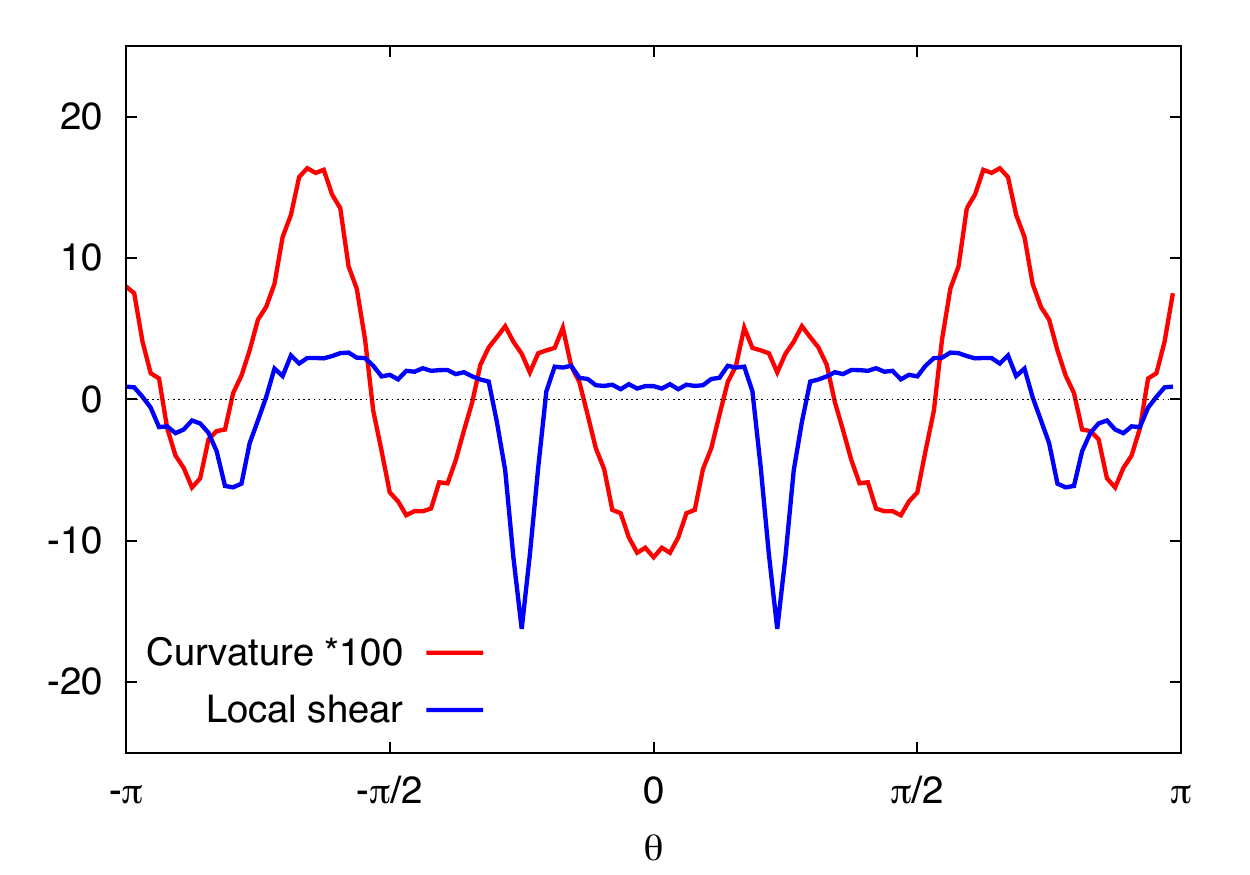}
\caption{Local shear and normal curvature (radial component of $\bkappa$) plotted as a function of poloidal angle $\theta$, on a fixed field line of W7-X.  The field line chosen, labeled $\alpha = 0$, passes through the outboard location of maximum bad curvature; this field line has very small global (average) magnetic shear, which is set to zero in the simulations for convenience.  Highly-localized ``spikes'' of magnetic shear are a generic feature of stellarator equilibria and occur where a magnetic surface bends sharply, \eg where field lines pass over a helical ridge.  The precise definitions of normal curvature and local shear used in these plots are given respectively by Eqns.~(148) and (152) of \citet{xanthopoulos-gist}.  Note that the local shear is essentially the derivative along the field line of the projection of $\bnabla\alpha$ in the radial direction.}
\label{shear-spike-fig}
\end{figure}

In a stellarator strong ``spikes'' of magnetic shear can appear at helical ridges, having the potential to trap ITG modes, in analogy to a quantum mechanical wavefunction being trapped in an infinite square well potential.  This local shear (termed ``local ripple shear'' by \citet{waltz-boozer}) can be strong, even in the absence of global shear.  Thus, we consider a model geometry that is uniform in an interval $[-\Lpar, \Lpar]$ and bounded by strong spikes of magnetic shear, as shown in Fig.~\ref{shear-spike-fig} for a field line passing through the region of maximum bad curvature.  The motivation behind this calculation is two-fold, namely (1) to explore the stabilizing influence of strong regions of local shear, and (2) to provide a simple fully-kinetic paradigm, beyond the slab ITG analysis, to describe the effect of localizing a mode along the magnetic field.  Some of the basic results of ballooning theory needed for this section are reviewed in Appx.~\ref{ballooning-appx}.  

Let us model the effect of the regions of strong local shear by taking

\begin{equation}
k_{\perp}(\ell) = \begin{cases} k_{\perp 0}, &\quad |\ell| \leq \Lpar, \\
\infty, & \quad |\ell| > \Lpar. \end{cases} \label{boxing-condition-eqn}
\end{equation}

\noindent This implies $\od = \infty$ and $J_0 = 0$ outside the boundaries, rendering the integrand in Eqn.~\ref{ballooning-disp-gen-eqn} zero for $|\ell^{\prime}| > \Lpar$.  This is because (1) $J_0' = 0$ and (2) the exponent ($\sim \exp(i M)$) becomes increasingly oscillatory at large $\od$, and so the velocity integral tends to zero.  The solutions we find will thus be bound states, effectively trapped in an infinite square well potential, which we call ``boxed'' modes.  However, the quantum mechanical analogy is not exact:  We are not solving a Schr\"{o}dinger equation, and our linear operator is not Hermitian.  Furthermore, the wave function in an infinite square well potential is pinned to zero at the boundaries, while our boundary conditions are those of ballooning theory.  In particular $\varphi$ is non-zero at the boundaries (though will decay rapidly outside the boundaries), while the distribution function itself need not even tend to zero outside the boundary as particles are free to stream out ballistically; however ballooning boundary conditions (Eqns.~\ref{bound-cond-a} and \ref{bound-cond-b}) require that none, as measured by the perturbed distribution $g$, can enter.  Note that, generally speaking, ``outgoing'' boundary conditions guarantee that free energy can only be lost at the boundaries.

Taking the integrand to be zero outside the domain in Eqn.~\ref{ballooning-disp-gen-eqn}, we set the bounds of the integral as $[-\Lpar, \Lpar]$. We then normalize the parallel coordinate $z = \ell/\Lpar$ and change to dimensionless frequencies and velocities, \ie $\Omega = \omega \Lpar/\vth$, $\Os = \os \Lpar/\vth$, \etc.  Then the integral dispersion relation\cite{cht-iii} for $\varphi$ and $\Omega$ (see Eqn.~\ref{ballooning-disp-gen-eqn}) can be expressed

\begin{equation}
\varphi(z) = \int_{-1}^{1}dz^{\prime}\varphi(z^{\prime}) \int_0^{\infty}\frac{d\xpar}{\xpar} \int_0^{\infty} x_{\perp}dx_{\perp}\exp\left(\frac{i}{\xpar}|z^{\prime}-z|(\Omega-\Odt)\right) H_*(\Omega),\label{full-integral-eqn}
\end{equation}

\noindent where

\begin{equation}
H_* = -\frac{2 i}{\sqrt{\pi}(1+\tau)} [\Omega - \Os(1 + \eta(x^2-3/2))]\exp(-x^2)J_0^2(x_{\perp}\sqrt{2b_0}),
\end{equation}

and $b_0 = k_{\perp 0}^2\rho^2$.  We solve Eqn.~\ref{full-integral-eqn} in two limits below:

\subsubsection{$\od$ = 0}

For $\od = 0$ we can evaluate the velocity integrals analytically and thereby arrive at a simpler equation.  Using Eqns.~\ref{Weber-integral-0}, \ref{Gn-def} and \ref{Weber-integral-1}, we find that Eqn.~\ref{full-integral-eqn} reduces to

\begin{equation}
\varphi(z) = \int_{-1}^{1} K(|z^{\prime} - z|, \O) \varphi(z^{\prime}) dz^{\prime}
\end{equation}

\noindent where

\begin{equation}
K(|z^{\prime} - z|, \O) = \frac{i/\sqrt{\pi}}{1 + \tau}\left[ \OsT\Gamma_0G_1(\Omega|z^{\prime}-z|) -((\Omega-\Os(1-\eta b -\eta/2))\Gamma_0 - \OsT b \Gamma_1)G_0(\Omega|z^{\prime}-z|)\right]
\end{equation}

\noindent and

\begin{equation}
G_n(y) = \int_0^{\infty} \exp(-x^2 + i y/x)\; x^{2n-1} dx,
\end{equation}

\noindent We solve this integral equation numerically by discretizing using a regularly spaced grid in $z$.  The function $G_0$ exhibits a logarithmic singularity at zero argument.  To handle this singularity numerically, we use a quadrature rule that is computed by interpolating $\varphi(z)$ using three points (\ie with a quadratic polynomial) and then integrating analytically.

\begin{figure}
\includegraphics[width=0.95\columnwidth]{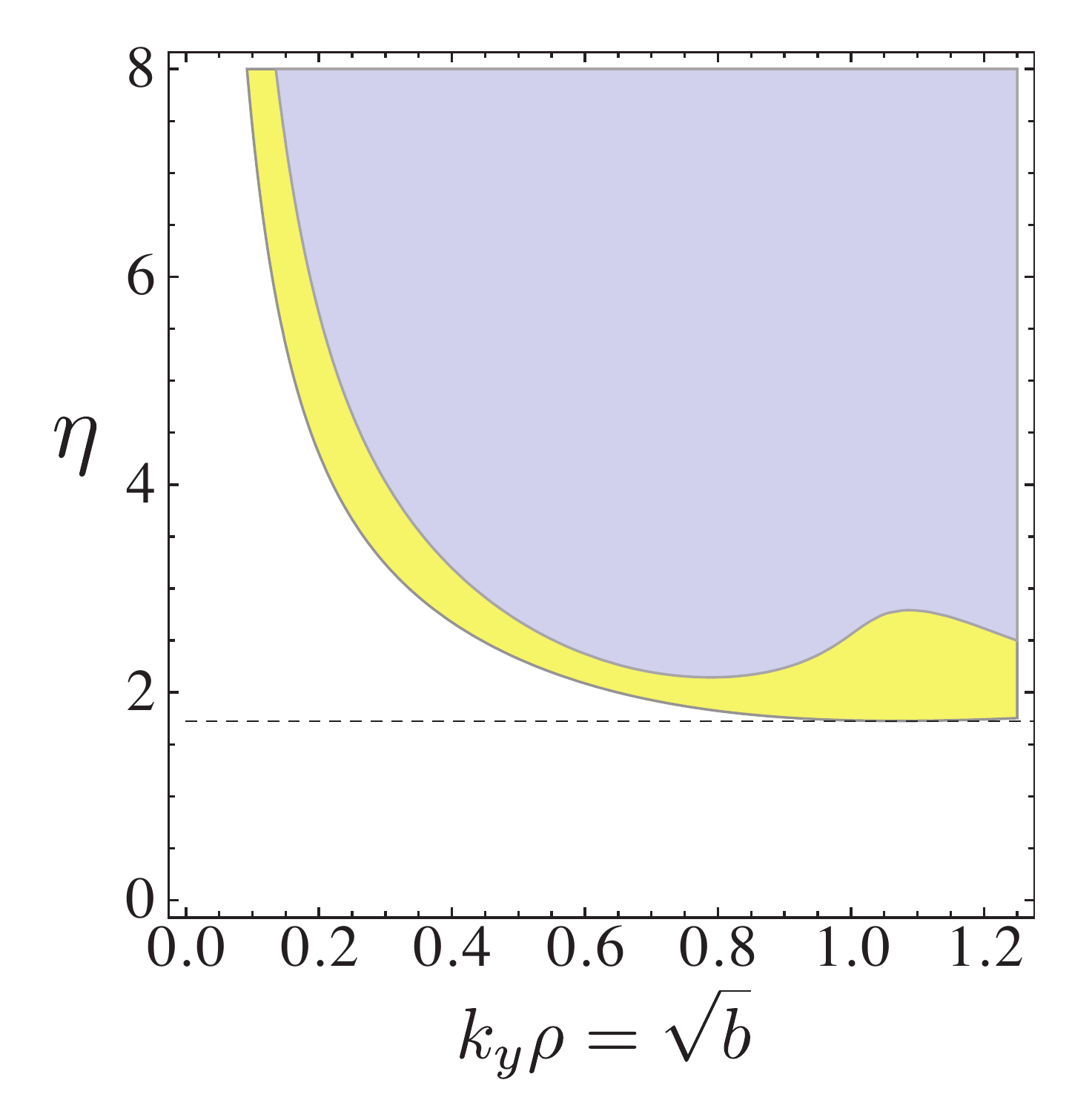}
\caption{Regions of instability:  The boxed mode (blue, $\Lpar/L_n = 7$) is compared with the ``fundamental'' slab mode (yellow, $\kpar = \pi/2\Lpar$, using Eqn.~\ref{slab-instability-criterion-eqn}).  Absolute minimum value of $\eta = 1.725$ indicated with dashed line.}
\label{boxed-marginal-fig}
\end{figure}

\begin{figure}
\includegraphics[width=0.95\columnwidth]{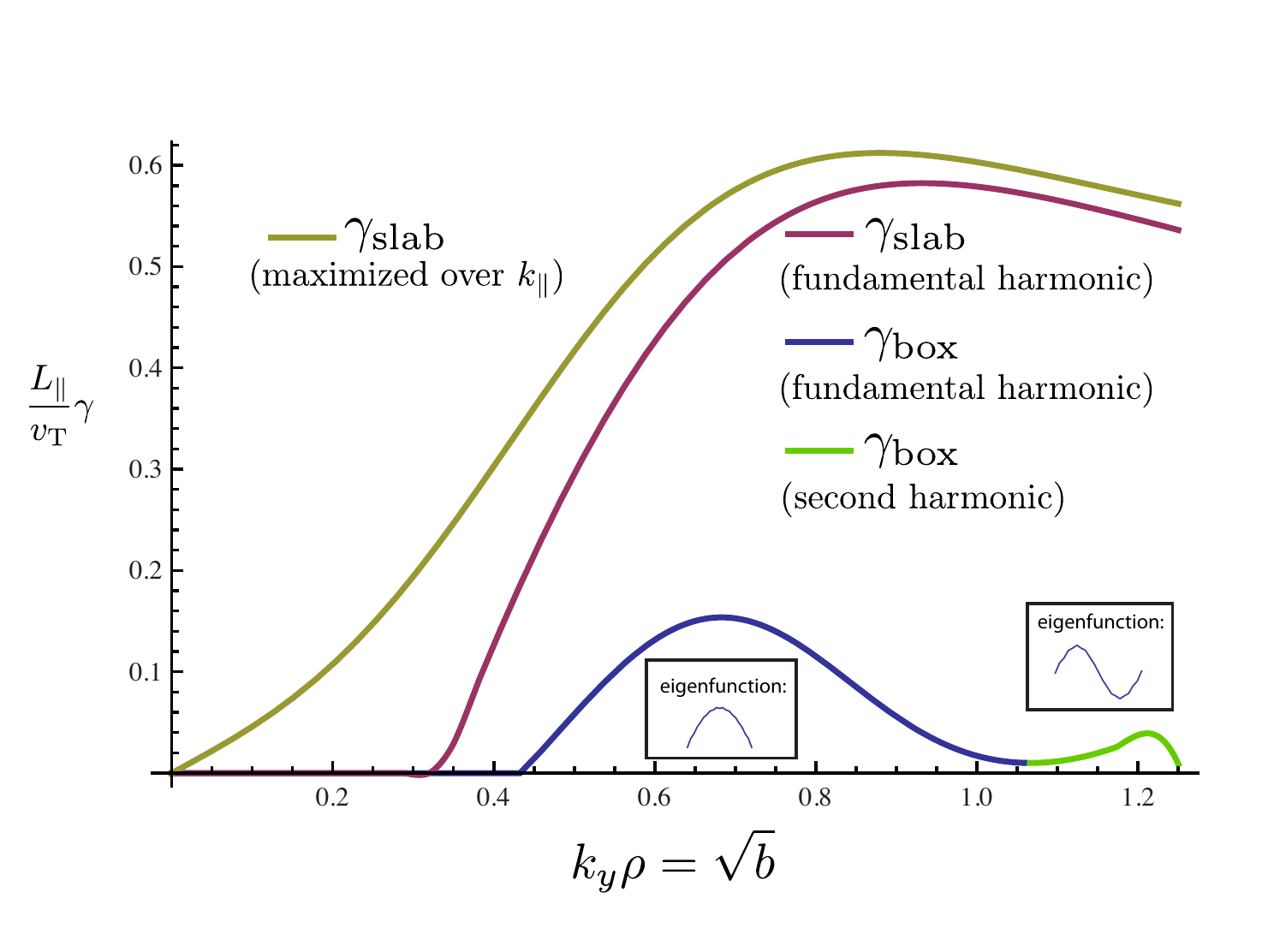}
\caption{Comparison of growth rates of slab and boxed modes for $\eta = 3$ and $\Lpar/L_n = 7$.  The ``fundamental'' slab mode has $\kpar = \pi/2\Lpar$.  For these parameters, two harmonics are found for the boxed mode; the transition between the two modes is indicated by a change in the plot color slightly above $k_y\rho = 1$.  Small inlays show the real part of eigenfunction for the two harmonics of the boxed mode.}
\label{boxed-slab-gamma-compare-fig}
\end{figure}

We find similar marginal stability contours as those depicted in Fig.~\ref{eta-b-stability-fig} by taking $\kpar = \pi/\Lpar$.   Fig.~\ref{boxed-marginal-fig} demonstrates this for the case of $\Lpar/L_n = 7$.  The growth rate of the boxed mode is generally less than slab values; see Fig.~\ref{boxed-slab-gamma-compare-fig}.

\subsubsection{$\od \neq 0$}\label{}

Solutions of the boxed mode indicate that the effect of the finite domain is stabilizing as compared with the unbound curvature-driven ITG mode ($\kpar = 0$); see Fig.~\ref{SAW7-X-vs-W7-X-fig} for an example using parameters relevant to W7-X ($\Lpar/\Reff = 0.96$, $\Reff/L_T = 31.6$ and $\eta = 3.0$).  This is generally consistent with transit-frequency damping, \ie the heuristic growth rate reduction rule $\gamma \rightarrow \gamma - \vth/2\Lpar$ provides a good estimate of the effect.

\subsubsection{Numerical experiments with localized shear spikes}

To investigate the stabilizing effect of shear spikes in realistic geometry, we have performed numerical experiments using the GENE code, in the flux tube domain.  A W7-X equilibrium is modified by removing the shear spikes (see Fig.~\ref{shear-spike-fig}).  This is done by direct manipulation of the metrics of the field-aligned coordinate system; see \citet{xanthopoulos-gist} for details.  The basic idea is to replace the local shear with a constant (along the field line), to mimic that of a circular tokamak.  This is achieved by first fixing the relationship between two of three metrics.  Then the third metric (proportional to $|\bnabla \alpha|^2$) is determined by requiring $B^2(\theta)$ to be preserved.  The magnetic drift frequency, however, is determined from the original W7-X equilibrium, so that the drift well contained within the shear spikes is preserved.  Thus, the procedure modifies only the value of $k_{\perp}$ that appears in the Bessel functions of the gyrokinetic system.

\begin{figure}
\includegraphics[width=0.95\columnwidth]{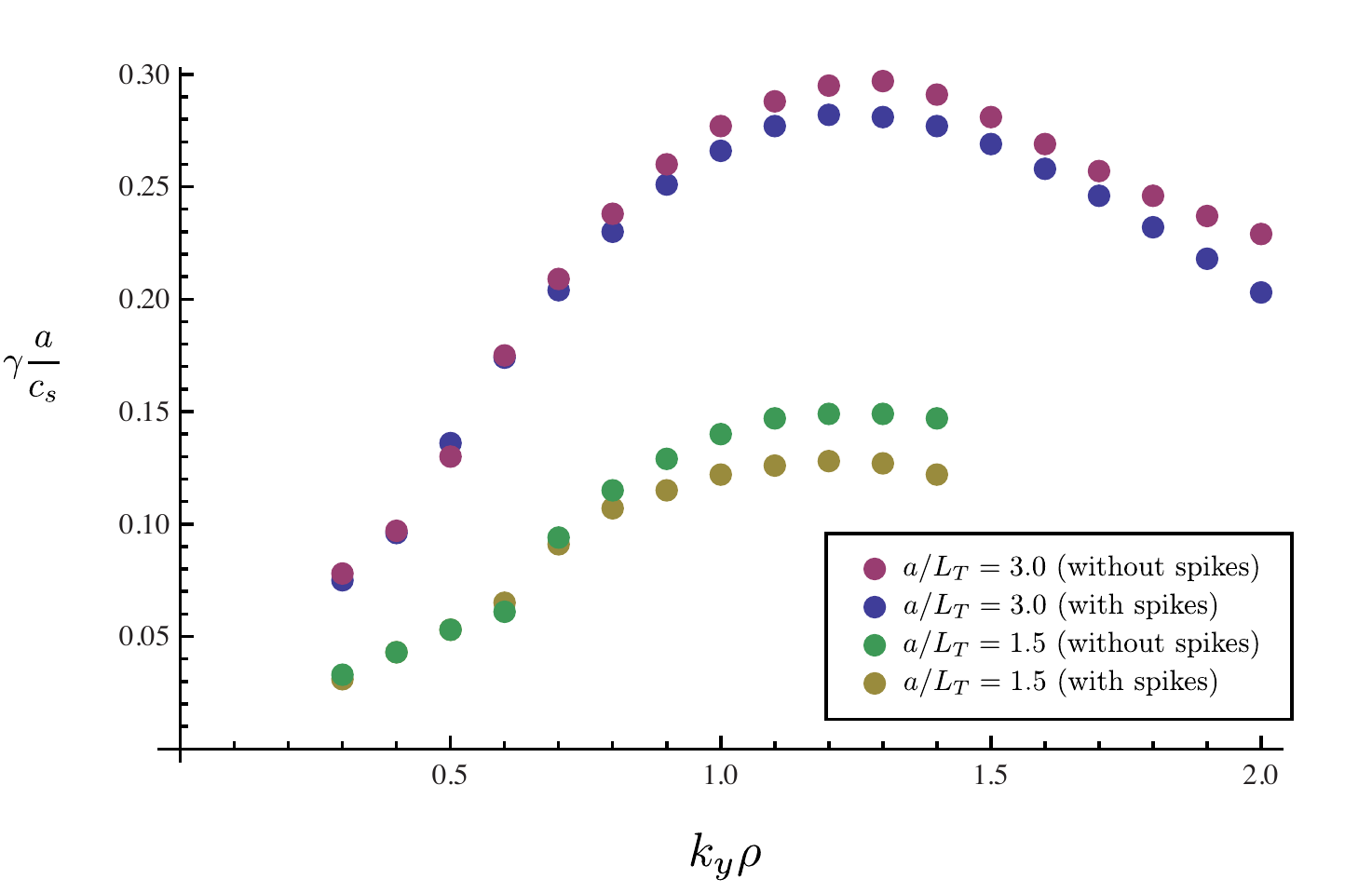}
\caption{Linear simulations in realistic geometry using GENE: ITG growth rates compared between W7-X equilibrium ($\alpha = 0$ flux tube) and a fictitious version that has the spikes of strong shear removed.  Two temperature gradients are shown, with similar results:  Localized shear can only significantly effect the mode at large $k_y$ ($k_{\alpha}$).}
\label{shear-spike-results-fig}
\end{figure}

This ``modified equilibrium'' is fictional (does not correspond to an actual MHD equilibrium), but the instability calculation that follows is mathematically well-defined.  The growth rates obtained are compared to those of the original W7-X equilibrium.  We find that the local shear only affects the growth rate at sufficiently large $k_y\rho$; see Fig.~\ref{shear-spike-results-fig}.  This is consistent with the analysis in Sec.~\ref{small-wt-sec}: FLR-induced shear suppression scales more strongly with $k_y$ than other effects in Eqn.~\ref{cht-nonres-eqn} and is thus negligible at low $k_y$ and dominant at large $k_y$.  The behavior shown in Fig.~\ref{shear-spike-results-fig} can also be anticipated from the fact that the amplification of $k_\perp$ (under the action of local shear) corresponds to a factor of about $5$-$6$ in W7-X; see Fig.~\ref{local-shear-vs-amp-fig}.  Thus, an approximate criterion for stabilization by ``boxing'' would be $k_y\rho A(\theta) > C$ (evaluated at the maximum of $A$), where $C$ is an order-unity constant, perhaps $5$.  However, as Fig.~\ref{eigenmode-localize-compare-fig} demonstrates, the local quadratic drift well, determined mostly by magnetic curvature, seems to mostly explain the localization of bound ITG modes in W7-X.

\section{Conclusion}

We summarize our findings as follows.  In line with conventional wisdom, we have found that the ITG mode is slab-like at long perpendicular wavelengths (see Secs.~\ref{gen-disp-sec} and \ref{slab-bloch-mode-sec}), while toroidally driven, localized modes are present at short wavelength.  If the local shear at a drift well is sufficiently small, the local dispersion relation for the toroidal mode can be obtained at short wavelengths as the mode becomes strongly ballooning and its frequency exceeds the transit frequency (see Sec.~\ref{small-wt-sec}); however, FLR-induced shear cannot be generally neglected in the strongly-ballooning limit and in that case the limit may not exist in which the local dispersion relation for the toroidal branch is valid.  The estimate $\kpar \sim 1/\Lpar$ is valid for determining the lower critical perpendicular wavenumber $\kyc$ (since the ITG mode is slab-like at long wavelengths).  It was also found in the study of the boxed ITG mode (Sec.~\ref{shear-well-sec}), that the growth rate is suppressed by localizing the ITG mode along a field line and the reduction of the growth rate seems consistent with the heuristic rule $\gamma \rightarrow \gamma - \vth/2\Lpar$.  This is because bound (localized) ITG modes radiate energy irreversibly at the boundaries.  Although the idealized model (assuming spikes of infinite shear) always shows stabilization of the ITG mode, the efficacy of ``boxing'' by localized shear in realistic geometries depends on a sufficiently high perpendicular wavenumber to activate FLR suppression; according to numerical evidence the effect is negligible in W7-X.  It is concluded that mode localization within a drift well of W7-X is determined by normal curvature.

Overall, our investigation supports the view that the ITG mode is well-characterized by local gradient scale lengths.  This implies that the ITG mode in W7-X will be more slab-like than that of conventional tokamak plasmas, simply by virtue of the fact that higher values of the instability parameter $\kappa$ can be accessed, allowing for a larger range of unstable modes for which the condition $\od \ll \op$ is satisfied; see the discussion in Sec.~\ref{gen-disp-sec}.  Generally, we also conclude that optimization of the ITG mode could be achieved by reducing the maximum magnitude of bad curvature and also reducing the connection lengths between regions of good and bad curvature.  Optimizing magnetic shear is another possibility: For W7-X, the rotational transform profile (and thus the accessible values of global shear) is significantly constrained (low-order rational surfaces are avoided and $\iota \sim 1$ at the edge).  However, in such cases one can still imagine preferentially distributing local magnetic shear so that it is negative around areas of bad curvature, as already occurs at the outboard triangular cross-section of W7-X.

\section{Acknowledgements}

We thank T. Bird, A. Bhattacharjee and G. Hammett for helpful discussions, suggestions and comments.  The GENE simulations were performed on the IFERC supercomputer.  This work was funded by the RCUK Energy Programme [grant number EP/I501045] and the European Communities under the contract of Association between EURATOM and CCFE. The views and opinions expressed herein do not necessarily reflect those of the European Commission.

\appendix

\section{Weber-type Integrals}\label{weber-int-appx}

For $p \in \field{R}$, $p > 0$ we have the Weber integral:

\begin{equation}
\int_0^{\infty} xdx \exp(-px^2)J_0^2(ax) = \frac{1}{2p} \exp(-a^2/2p)I_0(a^2/2p).
\end{equation}

\noindent For $p = 1$, we have

\begin{equation}
\int_0^{\infty} xdx \exp(-x^2)J_0^2(\sqrt{2b}x) = \frac{1}{2} \Gamma_0(b),\label{Weber-integral-0}
\end{equation}

\noindent where

\begin{equation}
\Gamma_n(b) = \exp(-b)I_n(b)\label{Gn-def}
\end{equation}

\noindent It follows for any integer $n$ that

\begin{equation}
\int_0^{\infty} x^{2n + 1}dx \exp(-px^2)J_0^2(ax) = (-1)^n\frac{d^n}{dp^n}\left(\frac{1}{2p} \exp(-a^2/2p)I_0(a^2/2p)\right).
\end{equation}

\noindent For $n = 1$ we have

\begin{equation}
\int_0^{\infty} x^3dx \exp(-x^2)J_0^2(\sqrt{2b}x) = \frac{1}{2}(1-b)\Gamma_0(b) + \frac{1}{2}b\Gamma_1(b).\label{Weber-integral-1}
\end{equation}

\section{Slab mode derivations}\label{slab-mode-appx}

The marginal stability criterion can be analytically derived from Eqn.~\ref{slab-disp-eqn} for the slab ITG mode.\cite{kadomtsev-pogutse}  By setting the real and imaginary parts separately to zero, two equations are obtained.   Let us take the limit $\Re[\xi] \rightarrow \xi_0$ and $\Im[\xi] \rightarrow 0+$ and note that the finite contribution to the imaginary part of the equation must come from $Z(\xi)$ (which can be evaluated by the Plemelj theorem) and thus one equation is obtained by setting the coefficient of $Z$ to zero and the other equation is formed by the remaining terms:

\begin{eqnarray}
0 = \Gamma_0\left[\xi_0 + \frac{\os}{\op}\left(\left(\frac{3\eta}{2} - 1\right) - \xi_0^2 \eta \right)\right] - \frac{\os}{\op} \eta ((1-b) \Gamma_0 + b \Gamma_1),\label{slab-crit-deriv-eqn-1}\\
0 = 1 + \tau - \Gamma_0\frac{\os}{\op}\xi_0 \eta,\label{slab-crit-deriv-eqn-2}
\end{eqnarray}

\noindent Eqn.~\ref{slab-crit-deriv-eqn-1} can be used to eliminate $\xi_0$ from Eqn.~\ref{slab-crit-deriv-eqn-1}, yielding a quadratic equation in $\eta$; the positive root of that equation then gives the critical $\eta$ of Eqn.~\ref{slab-instability-criterion-eqn}.  

To obtain the correction to $\xi$, let us define the limit more precisely.  We denote $\lim^{\prime}$ as the above limit, namely $\Re[\xi] \rightarrow \xi_0$ and $\Im[\xi] \rightarrow 0+$.  Now let us expand $\xi = \xi_0 + \xi_1$, where $\xi_1$ is a small correction with a positive imaginary part.  We expand the plasma dispersion function around $\xi_0$ as $Z(\xi) = Z_0 + Z_1$, where $Z_0 = \lim^{\prime} Z(\xi)$ and $Z_1 = \xi_1\lim^{\prime} Z'(\xi)$.  Now for fixed $\op$, we may choose a sufficiently large value of $\eta$ so that there exists a range of unstable wavenumbers $k_y$; see Fig.~\ref{eta-b-stability-fig}.  We expand about the lower critical $k_y = \kyc + \delta k_y$ and then order $\xi_1/\xi_0 \sim {\cal O} (\delta k_y/\kyc)$.  For simplicity, let us take $b = 0$ while holding $\os$ finite so that $\Gamma_0 = 1$ and $\Gamma_1 = 0$.  At zeroth order we have $\xi_0 = (1+\tau) \op/\osT$.  At first order we find a solution of the form $\xi_1 = F(\xi_0, \tau) \eta \delta \os/\op $.  For $\eta$ larger than the threshold value of $2$, it can be inferred from Eqn.~\ref{lw-criterion} that $\xi_0$ depends weakly on $\eta$ and tends to a constant for large $\eta$, in which case we obtain $\gamma = \op\Im[\xi_1]\propto \delta k_y\rho \vth/L_T$ as indicated in Fig.~\ref{local-disp-gamma-fig}.

\section{Low transit frequency limit}\label{small-wt-appx}

Eqn.~\ref{cht-eqn} may be derived by an asymptotic expansion of the linear gyrokinetic system in the limit $\ot/\omega \ll 1$.  We will need accuracy to two orders:

\begin{eqnarray}
g = g_0 + g_1 + g_2 + ... \\
\varphi = \varphi_0 + \varphi_1 + \varphi_2 + ...
\end{eqnarray}

\noindent Eqn.~\ref{gk-eqn} at each order can be solved iteratively for $g_n$ yielding

\begin{eqnarray}
g_0 = \left(\frac{\omega - \ost}{\omega - \od}\right) f_0 J_0 \varphi_0, \label{h0-eqn}\\
g_n = \left(\frac{-i \vpar}{\omega - \od}\frac{\partial}{\partial \ell}\right)^n g_0 + \left(\frac{\omega - \ost}{\omega - \od}\right) f_0 J_0 \varphi_n,\label{hn-eqn}
\end{eqnarray}



Using the expressions in Eqns.~\ref{h0-eqn} and \ref{hn-eqn} we substitute $g = g_0 + g_1 + g_2$ into Eqn.~\ref{qn-eqn}, whereby the first term in Eqn.~\ref{hn-eqn} integrates to zero by oddness in $\vpar$, and what remains is Eqn.~\ref{cht-eqn}.

\section{Ballooning theory}\label{ballooning-appx}

Let us return to the fully gyrokinetic treatment of the ITG mode.  Eqn.~\ref{gk-eqn} is differential at first order in $\ell$ and can be recast in integral form by taking the right-hand-side as a source and integrating directly, as shown by \citet{cht-iii}.  This is done for each sign $\sigma = \mbox{sign}(\vpar)$ separately, because we will need to use $\sigma$-dependent boundary conditions.  Rewriting Eqn.~\ref{gk-eqn} using $\vpar = \sigma|\vpar|$ we have

\begin{equation}
\frac{\partial g}{\partial \ell} - i\frac{\sigma}{|\vpar|}(\omega - \odt)g = -i\frac{\sigma}{|\vpar|}(\omega - \ost)J_0\varphi f_0\label{gk-eqn-2}
\end{equation}

\noindent We multiply this equation by the quantity $\exp(i\sigma M(\ell_0, \ell))$, where

\begin{equation}
M(\ell_0, \ell) = \int_{\ell_0}^{\ell} \frac{\omega - \odt(\ell^{\prime})}{|\vpar(\ell^{\prime})|} d\ell^{\prime},
\end{equation}

\noindent and integrate to arrive at the expression

\begin{equation}
g(\ell) = g_0 \exp(-i\sigma M(\ell_0, \ell)) - i\sigma(\omega - \ost)f_0\int_{\ell_0}^{\ell}\frac{J_0^{\prime}}{|\vpar^{\prime}|}\varphi(\ell^{\prime})\exp(i\sigma M(\ell^{\prime}, \ell))d\ell^{\prime},\label{g-soln-0}
\end{equation}

\noindent where $\vpar^{\prime} = \vpar(\ell^{\prime})$ and $J_0^{\prime} = J_0(k_{\perp}(\ell^{\prime})v_{\perp}(\ell^{\prime})/\Omega)$.

We must now decide what to do with the constant of integration $g_0$.  It is determined in ballooning theory by taking $\Im[\omega] = \gamma > 0$ and $\ell_0 = -\sigma \infty$, then noting that exponential divergence of the term $g_0 \exp(-i \sigma M)$ will occur unless $g_0 = 0$.  This is equivalent to applying the boundary conditions

\begin{eqnarray}
&g(\vpar > 0, \ell = -\infty) = 0,\label{bound-cond-a}\\ 
&g(\vpar < 0, \ell = \infty) = 0.\label{bound-cond-b}
\end{eqnarray}

\noindent These particular ``outgoing'' boundary conditions are also a statement of causality, \ie they state that no source of particles is fueling the instability from $\pm \infty$.  Such a mode would require communication across an infinite distance to coordinate the instability and this cannot occur in finite time, let alone on the timescales of the instability growth and saturation.  

Note that ballooning theory is necessary for modes which decay sufficiently slowly that periodicity is a nontrivial constraint, while decaying sufficiently fast that the inversion of the ballooning transformation does not diverge.  That is, for sufficiently localized modes (\ie modes that ``balloon'' strongly enough to have nearly-zero amplitude at $\theta = \pm \pi$) ballooning theory is not needed to describe it since there is no problem of enforcing periodicity.  In terms of the quantum mechanical analogy, we may say that if the mode does not significantly tunnel between drift wells, then we need only consider the local properties around a given well, as done in Sec.~\ref{small-wt-sec}.

Now applying outgoing boundary conditions, Eqn.~\ref{g-soln-0} becomes

\begin{equation}
g(\ell) = - i\sigma(\omega - \ost)f_0\int_{-\sigma \infty}^{\ell}\frac{J_0^{\prime}}{|\vpar^{\prime}|}\varphi(\ell^{\prime})\exp(i\sigma M(\ell^{\prime}, \ell))d\ell^{\prime}.\label{g-soln-1}
\end{equation}

\noindent Substituting this into Eqn.~\ref{qn-eqn} yields an embedded-eigenvalue integral equation for $\varphi(\ell)$ and eigenvalue $\omega$:

\begin{equation}
(1+ \tau)\varphi(\ell) = \frac{-2i}{\vth\sqrt{\pi}}\int_{0}^{\infty} \frac{d\xpar}{\xpar} \int_{0}^{\infty} \xperp d\xperp J_0 (\omega - \ost) \int_{-\infty}^{\infty} d\ell^{\prime} J_0^{\prime} \exp(-x^2 + i \sgn(\ell-\ell^{\prime})M(\ell^{\prime}, \ell))\varphi(\ell^{\prime}),\label{ballooning-disp-gen-eqn}
\end{equation}

\noindent where $\sgn$ gives the sign of its argument.  This is a simplified form of the general solution given by \citet{cht-iii} as it retains $\ell$-dependence only in $k_\perp$ and $\od$ and $\varphi$.

\section{On weak magnetic shear and periodicity}\label{weak-shear-disc-appx}

For the case of unbound (non-decaying or weakly decaying) modes in zero (or nearly-zero\footnote{Note that the limit of weak shear connected continuously with the zero shear case in Sec.~\ref{small-wt-sec} as localized modes become delocalized slab modes; see also \citet{candy-waltz-rosenbluth}.}) global magnetic shear, the ${\bf k}_{\perp}$ of Eqn.~\ref{kperp-eqn2} obeys periodicity in poloidal and toroidal angles ($\theta$, $\zeta$).  Observe that the non-periodic term of Eqn.~\ref{kperp-eqn2} is zero with zero global shear ($\iota' = d\iota/d\psi = 0$).  However the eikonal $\exp(i S)$ will generally not obey periodicity, except for special cases and so periodic solutions $g(\theta, \zeta)$ of Eqn.~\ref{gk-eqn} will not generally be periodic.  To overcome this problem, we modify the eikonal approach by taking

\begin{equation}
S = m\theta - n\zeta,
\end{equation}

\noindent where we have neglected radial variation of our mode, and $m$ and $n$ are integers chosen to be close to resonance so as to nearly satisfy ${\bf B}\cdot\bnabla S = 0$.  By writing $n/m = \iota + \delta\iota$, and repurposing the notation $k_{\alpha} = m$, we may write $S = k_{\alpha}(\alpha - \delta \iota \zeta)$.  Then the gyrokinetic equation is

\begin{equation}
i\vpar \frac{\partial g}{\partial \ell} + (\omega - \odt - \kappa_{\parallel}\vpar)g = \varphi J_0(\omega - \ost)f_0.\label{gk-zero-shear}
\end{equation}

\noindent where $\kappa_{\parallel}(\theta, \zeta) = \delta\iota k_{\alpha}\hat{\bf b}\cdot\bnabla\zeta$, and all other quantities are defined as before.  The only difference from Eqn.~\ref{gk-eqn} is in the appearance of a parallel drift due to the small deviation of ${\bf B}\cdot\bnabla S$ from zero.  This form of the gyrokinetic equation is periodic in both $\theta$ and $\zeta$ and one can formulate true global solutions as Bloch modes that obey this periodicity (\ie having zero Bloch wavenumber.\footnote{Note that the toroidal $2\pi/N$-periodicity for stellarators formally introduces the possibility of a nonzero Bloch wavenumber $k'_{\zeta} = n\pi/N$, with $n$ ranging $0$ to $N -1$, but this formal treatment is beyond the scope of the present work.})  

One can estimate the size of this drift by noting that $\delta \iota \lesssim 1/k_{\alpha}$ and so $\kappa_{\parallel} \vth \sim \vth \hat{\bf b}\cdot\bnabla\zeta \sim \vth/R$.  For a tokamak this could be a substantial effect since $\kpar \vth \sim 1/qR$.  In a stellarator of sufficiently high field periodicity it should be negligible because we have $\kappa_{\parallel}/\kpar \sim 1/N \ll 1$, where $N$ is the number of field periods.

\section{Nearly-slab Bloch Mode Derivations}\label{nearly-slab-appx}

Expanding in the small parameter $\od/\omega$, we first write the mode frequency as $\omega = \omega_0 + \omega_1 + \omega_2$, and by periodicity our solution must be of the form $g = g_0\exp(i M \theta) + g_1(\theta) + g_2(\theta)$, $\phi = \phi_0\exp(i M \theta) + \phi_1(\theta) + \phi_2(\theta)$.  The zeroth order solution is the slab mode of Sec.~\ref{slab-sec} with frequency $\omega_0$ and parallel wavenumber $\kpar = M \ot/\vth$.  By periodicity, the first order solution must be of the form $\phi_1 = \sum_n \phi_1^{(n)}\exp(i n \theta)$, $g_1 = \sum_n g_1^{(n)}\exp(i n \theta)$.  The equation for $n = M$ can be combined with quasi-neutrality, Eqn.~\ref{qn-eqn}, to show $\omega_1 = 0$.  Then the $n = M \pm m_d$ equations yield solutions for $g_1^{(M \pm m_d)}$ and $\phi_1^{(M \pm m_d)}$.  All other components of the first order solution are zero.  At second order, we again must have a series solution $\phi_2 = \sum_n \phi_2^{(n)}\exp(i n \theta)$, $g_2 = \sum_n g_2^{(n)}\exp(i n \theta)$, and the $n = M$ component of this equation yields Eqn.~\ref{o2-soln} with

\begin{equation}
{\mathcal I} = \frac{\frac{1}{2}\int d^3{\bf x} \hat{f}_0 J_0^2 \displaystyle\sum_{n=M \pm m_d} \frac{(\xpar^2 + \xperp^2/2)(\omega_0 - \ost)}{(\omega_0 - n\ot\xpar)(\omega_0 - M \ot\xpar)}\left\{ \frac{1}{2}\frac{\xpar^2 + \xperp^2/2}{\omega_0 - M\ot\xpar} + r_n\right\}}{\int d^3{\bf x} \hat{f}_0 J_0^2 \frac{\omega_0 - \ost}{(\omega_0 - M\ot\xpar)^2}},\label{o2-int}
\end{equation}

\noindent where we define $d^3{\bf x} = 2\pi \xperp d \xperp d\xpar$ and we introduce the definitions

\begin{equation}
\hat{f}_0 = \vth^3 f_0/n_0,
\end{equation}

\noindent and

\begin{equation}
r_n = \frac{1}{2}\frac{\int d^3{\bf x} \hat{f}_0 J_0^2[\xpar^2+\xperp^2/2]\frac{\omega_0 - \ost}{\omega_0 - M\ot\xpar}}{1 + \tau - \int d^3{\bf x} J_0^2 \hat{f}_0 \frac{\omega_0 - \ost}{\omega_0 - n \ot \xpar}}
\end{equation}


\bibliography{ITG-mode}

\begin{thebibliography}{28}%
\makeatletter
\providecommand \@ifxundefined [1]{%
 \@ifx{#1\undefined}
}%
\providecommand \@ifnum [1]{%
 \ifnum #1\expandafter \@firstoftwo
 \else \expandafter \@secondoftwo
 \fi
}%
\providecommand \@ifx [1]{%
 \ifx #1\expandafter \@firstoftwo
 \else \expandafter \@secondoftwo
 \fi
}%
\providecommand \natexlab [1]{#1}%
\providecommand \enquote  [1]{``#1''}%
\providecommand \bibnamefont  [1]{#1}%
\providecommand \bibfnamefont [1]{#1}%
\providecommand \citenamefont [1]{#1}%
\providecommand \href@noop [0]{\@secondoftwo}%
\providecommand \href [0]{\begingroup \@sanitize@url \@href}%
\providecommand \@href[1]{\@@startlink{#1}\@@href}%
\providecommand \@@href[1]{\endgroup#1\@@endlink}%
\providecommand \@sanitize@url [0]{\catcode `\\12\catcode `\$12\catcode
  `\&12\catcode `\#12\catcode `\^12\catcode `\_12\catcode `\%12\relax}%
\providecommand \@@startlink[1]{}%
\providecommand \@@endlink[0]{}%
\providecommand \url  [0]{\begingroup\@sanitize@url \@url }%
\providecommand \@url [1]{\endgroup\@href {#1}{\urlprefix }}%
\providecommand \urlprefix  [0]{URL }%
\providecommand \Eprint [0]{\href }%
\providecommand \doibase [0]{http://dx.doi.org/}%
\providecommand \selectlanguage [0]{\@gobble}%
\providecommand \bibinfo  [0]{\@secondoftwo}%
\providecommand \bibfield  [0]{\@secondoftwo}%
\providecommand \translation [1]{[#1]}%
\providecommand \BibitemOpen [0]{}%
\providecommand \bibitemStop [0]{}%
\providecommand \bibitemNoStop [0]{.\EOS\space}%
\providecommand \EOS [0]{\spacefactor3000\relax}%
\providecommand \BibitemShut  [1]{\csname bibitem#1\endcsname}%
\let\auto@bib@innerbib\@empty
\bibitem [{\citenamefont {Mynick}, \citenamefont {Pomphrey},\ and\
  \citenamefont {Xanthopoulos}(2010)}]{mynick-stelopt}%
  \BibitemOpen
  \bibfield  {author} {\bibinfo {author} {\bibfnamefont {H.~E.}\ \bibnamefont
  {Mynick}}, \bibinfo {author} {\bibfnamefont {N.}~\bibnamefont {Pomphrey}}, \
  and\ \bibinfo {author} {\bibfnamefont {P.}~\bibnamefont {Xanthopoulos}},\
  }\href {\doibase 10.1103/PhysRevLett.105.095004} {\bibfield  {journal}
  {\bibinfo  {journal} {Phys. Rev. Lett.}\ }\textbf {\bibinfo {volume} {105}},\
  \bibinfo {pages} {095004} (\bibinfo {year} {2010})}\BibitemShut {NoStop}%
\bibitem [{\citenamefont {Proll}\ \emph {et~al.}(2012)\citenamefont {Proll},
  \citenamefont {Helander}, \citenamefont {Connor},\ and\ \citenamefont
  {Plunk}}]{proll-prl}%
  \BibitemOpen
  \bibfield  {author} {\bibinfo {author} {\bibfnamefont {J.~H.~E.}\
  \bibnamefont {Proll}}, \bibinfo {author} {\bibfnamefont {P.}~\bibnamefont
  {Helander}}, \bibinfo {author} {\bibfnamefont {J.~W.}\ \bibnamefont
  {Connor}}, \ and\ \bibinfo {author} {\bibfnamefont {G.~G.}\ \bibnamefont
  {Plunk}},\ }\href {\doibase 10.1103/PhysRevLett.108.245002} {\bibfield
  {journal} {\bibinfo  {journal} {Phys. Rev. Lett.}\ }\textbf {\bibinfo
  {volume} {108}},\ \bibinfo {pages} {245002} (\bibinfo {year}
  {2012})}\BibitemShut {NoStop}%
\bibitem [{\citenamefont {Helander}, \citenamefont {Proll},\ and\ \citenamefont
  {Plunk}(2013)}]{resilience-pop-I}%
  \BibitemOpen
  \bibfield  {author} {\bibinfo {author} {\bibfnamefont {P.}~\bibnamefont
  {Helander}}, \bibinfo {author} {\bibfnamefont {J.~H.~E.}\ \bibnamefont
  {Proll}}, \ and\ \bibinfo {author} {\bibfnamefont {G.~G.}\ \bibnamefont
  {Plunk}},\ }\href@noop {} {\bibfield  {journal} {\bibinfo  {journal} {Phys.
  Plasmas}\ } (\bibinfo {year} {{\em accepted} 2013})}\BibitemShut {NoStop}%
\bibitem [{\citenamefont {Proll}, \citenamefont {Xanthopoulos},\ and\
  \citenamefont {Helander}(2013)}]{resilience-pop-II}%
  \BibitemOpen
  \bibfield  {author} {\bibinfo {author} {\bibfnamefont {J.~H.~E.}\
  \bibnamefont {Proll}}, \bibinfo {author} {\bibfnamefont {P.}~\bibnamefont
  {Xanthopoulos}}, \ and\ \bibinfo {author} {\bibfnamefont {P.}~\bibnamefont
  {Helander}},\ }\href@noop {} {\bibfield  {journal} {\bibinfo  {journal}
  {Phys. Plasmas}\ } (\bibinfo {year} {{\em accepted} 2013})}\BibitemShut
  {NoStop}%
\bibitem [{\citenamefont {Helander}\ \emph {et~al.}(2012)\citenamefont
  {Helander}, \citenamefont {Beidler}, \citenamefont {Bird}, \citenamefont
  {Drevlak}, \citenamefont {Feng}, \citenamefont {Hatzky}, \citenamefont
  {Jenko}, \citenamefont {Kleiber}, \citenamefont {Proll}, \citenamefont
  {Turkin},\ and\ \citenamefont {Xanthopoulos}}]{helander-a-comparison}%
  \BibitemOpen
  \bibfield  {author} {\bibinfo {author} {\bibfnamefont {P.}~\bibnamefont
  {Helander}}, \bibinfo {author} {\bibfnamefont {C.~D.}\ \bibnamefont
  {Beidler}}, \bibinfo {author} {\bibfnamefont {T.~M.}\ \bibnamefont {Bird}},
  \bibinfo {author} {\bibfnamefont {M.}~\bibnamefont {Drevlak}}, \bibinfo
  {author} {\bibfnamefont {Y.}~\bibnamefont {Feng}}, \bibinfo {author}
  {\bibfnamefont {R.}~\bibnamefont {Hatzky}}, \bibinfo {author} {\bibfnamefont
  {F.}~\bibnamefont {Jenko}}, \bibinfo {author} {\bibfnamefont
  {R.}~\bibnamefont {Kleiber}}, \bibinfo {author} {\bibfnamefont {J.~H.~E.}\
  \bibnamefont {Proll}}, \bibinfo {author} {\bibfnamefont {Y.}~\bibnamefont
  {Turkin}}, \ and\ \bibinfo {author} {\bibfnamefont {P.}~\bibnamefont
  {Xanthopoulos}},\ }\href {http://stacks.iop.org/0741-3335/54/i=12/a=124009}
  {\bibfield  {journal} {\bibinfo  {journal} {Plasma Phys. Control. Fusion}\
  }\textbf {\bibinfo {volume} {54}},\ \bibinfo {pages} {124009} (\bibinfo
  {year} {2012})}\BibitemShut {NoStop}%
\bibitem [{Note1()}]{Note1}%
  \BibitemOpen
  \bibinfo {note} {Note that the eikonal representation is only appropriate
  when variation of the mode occurs on a smaller scale than that of equilibrium
  quantities like magnetic curvature and shear. Although it is unknown to what
  extent this is satisfied in the radial direction, recent full-flux-surface
  simulations have revealed a loss of scale separation in the binormal
  direction for modes with small $k_\perp $. However, details of this will be
  left for a future paper.}\BibitemShut {Stop}%
\bibitem [{\citenamefont {Roberts}\ and\ \citenamefont
  {Taylor}(1965)}]{roberts-taylor}%
  \BibitemOpen
  \bibfield  {author} {\bibinfo {author} {\bibfnamefont {K.~V.}\ \bibnamefont
  {Roberts}}\ and\ \bibinfo {author} {\bibfnamefont {J.~B.}\ \bibnamefont
  {Taylor}},\ }\href {\doibase 10.1063/1.1761225} {\bibfield  {journal}
  {\bibinfo  {journal} {Physics of Fluids}\ }\textbf {\bibinfo {volume} {8}},\
  \bibinfo {pages} {315} (\bibinfo {year} {1965})}\BibitemShut {NoStop}%
\bibitem [{\citenamefont {Connor}, \citenamefont {Hastie},\ and\ \citenamefont
  {Taylor}(1980)}]{cht-iii}%
  \BibitemOpen
  \bibfield  {author} {\bibinfo {author} {\bibfnamefont {J.~W.}\ \bibnamefont
  {Connor}}, \bibinfo {author} {\bibfnamefont {R.~J.}\ \bibnamefont {Hastie}},
  \ and\ \bibinfo {author} {\bibfnamefont {J.~B.}\ \bibnamefont {Taylor}},\
  }\href {http://stacks.iop.org/0032-1028/22/i=7/a=013} {\bibfield  {journal}
  {\bibinfo  {journal} {Plasma Physics}\ }\textbf {\bibinfo {volume} {22}},\
  \bibinfo {pages} {757} (\bibinfo {year} {1980})}\BibitemShut {NoStop}%
\bibitem [{\citenamefont {Dewar}\ and\ \citenamefont
  {Glasser}(1983)}]{dewar-glasser}%
  \BibitemOpen
  \bibfield  {author} {\bibinfo {author} {\bibfnamefont {R.~L.}\ \bibnamefont
  {Dewar}}\ and\ \bibinfo {author} {\bibfnamefont {A.~H.}\ \bibnamefont
  {Glasser}},\ }\href {\doibase 10.1063/1.864028} {\bibfield  {journal}
  {\bibinfo  {journal} {Physics of Fluids}\ }\textbf {\bibinfo {volume} {26}},\
  \bibinfo {pages} {3038} (\bibinfo {year} {1983})}\BibitemShut {NoStop}%
\bibitem [{\citenamefont {Kim}\ and\ \citenamefont
  {Wakatani}(1995)}]{kim-wakatani}%
  \BibitemOpen
  \bibfield  {author} {\bibinfo {author} {\bibfnamefont {J.~Y.}\ \bibnamefont
  {Kim}}\ and\ \bibinfo {author} {\bibfnamefont {M.}~\bibnamefont {Wakatani}},\
  }\href {\doibase 10.1063/1.871407} {\bibfield  {journal} {\bibinfo  {journal}
  {Physics of Plasmas}\ }\textbf {\bibinfo {volume} {2}},\ \bibinfo {pages}
  {1012} (\bibinfo {year} {1995})}\BibitemShut {NoStop}%
\bibitem [{\citenamefont {Choi}\ and\ \citenamefont
  {Horton}(1980)}]{choi-horton}%
  \BibitemOpen
  \bibfield  {author} {\bibinfo {author} {\bibfnamefont {D.-I.}\ \bibnamefont
  {Choi}}\ and\ \bibinfo {author} {\bibfnamefont {W.}~\bibnamefont {Horton}},\
  }\href {\doibase 10.1063/1.862980} {\bibfield  {journal} {\bibinfo  {journal}
  {Physics of Fluids}\ }\textbf {\bibinfo {volume} {23}},\ \bibinfo {pages}
  {356} (\bibinfo {year} {1980})}\BibitemShut {NoStop}%
\bibitem [{\citenamefont {T.~M.~Antonsen}\ \emph {et~al.}(1996)\citenamefont
  {T.~M.~Antonsen}, \citenamefont {Drake}, \citenamefont {Guzdar},
  \citenamefont {Hassam}, \citenamefont {Lau}, \citenamefont {Liu},\ and\
  \citenamefont {Novakovskii}}]{antonsen}%
  \BibitemOpen
  \bibfield  {author} {\bibinfo {author} {\bibfnamefont {J.}~\bibnamefont
  {T.~M.~Antonsen}}, \bibinfo {author} {\bibfnamefont {J.~F.}\ \bibnamefont
  {Drake}}, \bibinfo {author} {\bibfnamefont {P.~N.}\ \bibnamefont {Guzdar}},
  \bibinfo {author} {\bibfnamefont {A.~B.}\ \bibnamefont {Hassam}}, \bibinfo
  {author} {\bibfnamefont {Y.~T.}\ \bibnamefont {Lau}}, \bibinfo {author}
  {\bibfnamefont {C.~S.}\ \bibnamefont {Liu}}, \ and\ \bibinfo {author}
  {\bibfnamefont {S.~V.}\ \bibnamefont {Novakovskii}},\ }\href {\doibase
  10.1063/1.871928} {\bibfield  {journal} {\bibinfo  {journal} {Physics of
  Plasmas}\ }\textbf {\bibinfo {volume} {3}},\ \bibinfo {pages} {2221}
  (\bibinfo {year} {1996})}\BibitemShut {NoStop}%
\bibitem [{\citenamefont {Kadomtsev}\ and\ \citenamefont
  {Pogutse}(1970)}]{kadomtsev-pogutse}%
  \BibitemOpen
  \bibfield  {author} {\bibinfo {author} {\bibfnamefont {B.~B.}\ \bibnamefont
  {Kadomtsev}}\ and\ \bibinfo {author} {\bibfnamefont {O.~P.}\ \bibnamefont
  {Pogutse}},\ }\href@noop {} {\bibfield  {journal} {\bibinfo  {journal} {Rev.
  Plasmas Phys.}\ }\textbf {\bibinfo {volume} {5}},\ \bibinfo {pages} {249}
  (\bibinfo {year} {1970})}\BibitemShut {NoStop}%
\bibitem [{\citenamefont {Biglari}, \citenamefont {Diamond},\ and\
  \citenamefont {Rosenbluth}(1989)}]{biglari}%
  \BibitemOpen
  \bibfield  {author} {\bibinfo {author} {\bibfnamefont {H.}~\bibnamefont
  {Biglari}}, \bibinfo {author} {\bibfnamefont {P.~H.}\ \bibnamefont
  {Diamond}}, \ and\ \bibinfo {author} {\bibfnamefont {M.~N.}\ \bibnamefont
  {Rosenbluth}},\ }\href {\doibase 10.1063/1.859206} {\bibfield  {journal}
  {\bibinfo  {journal} {Physics of Fluids B: Plasma Physics}\ }\textbf
  {\bibinfo {volume} {1}},\ \bibinfo {pages} {109} (\bibinfo {year}
  {1989})}\BibitemShut {NoStop}%
\bibitem [{\citenamefont {Cowley}, \citenamefont {Kulsrud},\ and\ \citenamefont
  {Sudan}(1991)}]{cowley-kulsrud}%
  \BibitemOpen
  \bibfield  {author} {\bibinfo {author} {\bibfnamefont {S.~C.}\ \bibnamefont
  {Cowley}}, \bibinfo {author} {\bibfnamefont {R.~M.}\ \bibnamefont {Kulsrud}},
  \ and\ \bibinfo {author} {\bibfnamefont {R.}~\bibnamefont {Sudan}},\
  }\href@noop {} {\bibfield  {journal} {\bibinfo  {journal} {Phys. Fluids B}\
  }\textbf {\bibinfo {volume} {3}},\ \bibinfo {pages} {2767} (\bibinfo {year}
  {1991})}\BibitemShut {NoStop}%
\bibitem [{\citenamefont {Chowdhury}\ \emph {et~al.}(2012)\citenamefont
  {Chowdhury}, \citenamefont {Brunner}, \citenamefont {Ganesh}, \citenamefont
  {Lapillonne}, \citenamefont {Villard},\ and\ \citenamefont {Jenko}}]{switg}%
  \BibitemOpen
  \bibfield  {author} {\bibinfo {author} {\bibfnamefont {J.}~\bibnamefont
  {Chowdhury}}, \bibinfo {author} {\bibfnamefont {S.}~\bibnamefont {Brunner}},
  \bibinfo {author} {\bibfnamefont {R.}~\bibnamefont {Ganesh}}, \bibinfo
  {author} {\bibfnamefont {X.}~\bibnamefont {Lapillonne}}, \bibinfo {author}
  {\bibfnamefont {L.}~\bibnamefont {Villard}}, \ and\ \bibinfo {author}
  {\bibfnamefont {F.}~\bibnamefont {Jenko}},\ }\href {\doibase
  10.1063/1.4759458} {\bibfield  {journal} {\bibinfo  {journal} {Physics of
  Plasmas}\ }\textbf {\bibinfo {volume} {19}},\ \bibinfo {eid} {102508}
  (\bibinfo {year} {2012})}\BibitemShut {NoStop}%
\bibitem [{\citenamefont {T.~M.~Antonsen}\ and\ \citenamefont
  {Lane}(1980)}]{antonsen-lane}%
  \BibitemOpen
  \bibfield  {author} {\bibinfo {author} {\bibfnamefont {J.}~\bibnamefont
  {T.~M.~Antonsen}}\ and\ \bibinfo {author} {\bibfnamefont {B.}~\bibnamefont
  {Lane}},\ }\href {\doibase 10.1063/1.863121} {\bibfield  {journal} {\bibinfo
  {journal} {Physics of Fluids}\ }\textbf {\bibinfo {volume} {23}},\ \bibinfo
  {pages} {1205} (\bibinfo {year} {1980})}\BibitemShut {NoStop}%
\bibitem [{\citenamefont {Wendell~Horton}, \citenamefont {Choi},\ and\
  \citenamefont {Tang}(1981)}]{horton-choi-tang}%
  \BibitemOpen
  \bibfield  {author} {\bibinfo {author} {\bibfnamefont {J.}~\bibnamefont
  {Wendell~Horton}}, \bibinfo {author} {\bibfnamefont {D.-I.}\ \bibnamefont
  {Choi}}, \ and\ \bibinfo {author} {\bibfnamefont {W.~M.}\ \bibnamefont
  {Tang}},\ }\href {\doibase 10.1063/1.863486} {\bibfield  {journal} {\bibinfo
  {journal} {Physics of Fluids}\ }\textbf {\bibinfo {volume} {24}},\ \bibinfo
  {pages} {1077} (\bibinfo {year} {1981})}\BibitemShut {NoStop}%
\bibitem [{\citenamefont {Romanelli}(1989)}]{romanelli-1989}%
  \BibitemOpen
  \bibfield  {author} {\bibinfo {author} {\bibfnamefont {F.}~\bibnamefont
  {Romanelli}},\ }\href {\doibase 10.1063/1.859023} {\bibfield  {journal}
  {\bibinfo  {journal} {Physics of Fluids B: Plasma Physics}\ }\textbf
  {\bibinfo {volume} {1}},\ \bibinfo {pages} {1018} (\bibinfo {year}
  {1989})}\BibitemShut {NoStop}%
\bibitem [{\citenamefont {Romanelli}, \citenamefont {Chen},\ and\ \citenamefont
  {Briguglio}(1991)}]{romanelli-chen-1991}%
  \BibitemOpen
  \bibfield  {author} {\bibinfo {author} {\bibfnamefont {F.}~\bibnamefont
  {Romanelli}}, \bibinfo {author} {\bibfnamefont {L.}~\bibnamefont {Chen}}, \
  and\ \bibinfo {author} {\bibfnamefont {S.}~\bibnamefont {Briguglio}},\ }\href
  {\doibase http://dx.doi.org/10.1063/1.859960} {\bibfield  {journal} {\bibinfo
   {journal} {Physics of Fluids B: Plasma Physics (1989-1993)}\ }\textbf
  {\bibinfo {volume} {3}},\ \bibinfo {pages} {2496} (\bibinfo {year}
  {1991})}\BibitemShut {NoStop}%
\bibitem [{\citenamefont {Bhattacharjee}\ \emph {et~al.}(1983)\citenamefont
  {Bhattacharjee}, \citenamefont {Sedlak}, \citenamefont {Similon},
  \citenamefont {Rosenbluth},\ and\ \citenamefont {Ross}}]{bhattacharjee}%
  \BibitemOpen
  \bibfield  {author} {\bibinfo {author} {\bibfnamefont {A.}~\bibnamefont
  {Bhattacharjee}}, \bibinfo {author} {\bibfnamefont {J.~E.}\ \bibnamefont
  {Sedlak}}, \bibinfo {author} {\bibfnamefont {P.~L.}\ \bibnamefont {Similon}},
  \bibinfo {author} {\bibfnamefont {M.~N.}\ \bibnamefont {Rosenbluth}}, \ and\
  \bibinfo {author} {\bibfnamefont {D.~W.}\ \bibnamefont {Ross}},\ }\href
  {\doibase 10.1063/1.864229} {\bibfield  {journal} {\bibinfo  {journal}
  {Physics of Fluids}\ }\textbf {\bibinfo {volume} {26}},\ \bibinfo {pages}
  {880} (\bibinfo {year} {1983})}\BibitemShut {NoStop}%
\bibitem [{\citenamefont {Candy}, \citenamefont {Waltz},\ and\ \citenamefont
  {Rosenbluth}(2004)}]{candy-waltz-rosenbluth}%
  \BibitemOpen
  \bibfield  {author} {\bibinfo {author} {\bibfnamefont {J.}~\bibnamefont
  {Candy}}, \bibinfo {author} {\bibfnamefont {R.~E.}\ \bibnamefont {Waltz}}, \
  and\ \bibinfo {author} {\bibfnamefont {M.~N.}\ \bibnamefont {Rosenbluth}},\
  }\href {\doibase 10.1063/1.1689967} {\bibfield  {journal} {\bibinfo
  {journal} {Physics of Plasmas}\ }\textbf {\bibinfo {volume} {11}},\ \bibinfo
  {pages} {1879} (\bibinfo {year} {2004})}\BibitemShut {NoStop}%
\bibitem [{\citenamefont {McMillan}\ and\ \citenamefont
  {Dewar}(2006)}]{mcmillan-dewar}%
  \BibitemOpen
  \bibfield  {author} {\bibinfo {author} {\bibfnamefont {B.}~\bibnamefont
  {McMillan}}\ and\ \bibinfo {author} {\bibfnamefont {R.}~\bibnamefont
  {Dewar}},\ }\href {http://stacks.iop.org/0029-5515/46/i=4/a=008} {\bibfield
  {journal} {\bibinfo  {journal} {Nuclear Fusion}\ }\textbf {\bibinfo {volume}
  {46}},\ \bibinfo {pages} {477} (\bibinfo {year} {2006})}\BibitemShut
  {NoStop}%
\bibitem [{\citenamefont {Connor}\ and\ \citenamefont
  {Hastie}(2004)}]{connor-hastie-2004}%
  \BibitemOpen
  \bibfield  {author} {\bibinfo {author} {\bibfnamefont {J.~W.}\ \bibnamefont
  {Connor}}\ and\ \bibinfo {author} {\bibfnamefont {R.~J.}\ \bibnamefont
  {Hastie}},\ }\href {http://stacks.iop.org/0741-3335/46/i=10/a=001} {\bibfield
   {journal} {\bibinfo  {journal} {Plasma Physics and Controlled Fusion}\
  }\textbf {\bibinfo {volume} {46}},\ \bibinfo {pages} {1501} (\bibinfo {year}
  {2004})}\BibitemShut {NoStop}%
\bibitem [{\citenamefont {Waltz}\ and\ \citenamefont
  {Boozer}(1993)}]{waltz-boozer}%
  \BibitemOpen
  \bibfield  {author} {\bibinfo {author} {\bibfnamefont {R.~E.}\ \bibnamefont
  {Waltz}}\ and\ \bibinfo {author} {\bibfnamefont {A.~H.}\ \bibnamefont
  {Boozer}},\ }\href {\doibase 10.1063/1.860754} {\bibfield  {journal}
  {\bibinfo  {journal} {Physics of Fluids B: Plasma Physics}\ }\textbf
  {\bibinfo {volume} {5}},\ \bibinfo {pages} {2201} (\bibinfo {year}
  {1993})}\BibitemShut {NoStop}%
\bibitem [{\citenamefont {Xanthopoulos}\ \emph {et~al.}(2009)\citenamefont
  {Xanthopoulos}, \citenamefont {Cooper}, \citenamefont {Jenko}, \citenamefont
  {Turkin}, \citenamefont {Runov},\ and\ \citenamefont
  {Geiger}}]{xanthopoulos-gist}%
  \BibitemOpen
  \bibfield  {author} {\bibinfo {author} {\bibfnamefont {P.}~\bibnamefont
  {Xanthopoulos}}, \bibinfo {author} {\bibfnamefont {W.~A.}\ \bibnamefont
  {Cooper}}, \bibinfo {author} {\bibfnamefont {F.}~\bibnamefont {Jenko}},
  \bibinfo {author} {\bibfnamefont {Y.}~\bibnamefont {Turkin}}, \bibinfo
  {author} {\bibfnamefont {A.}~\bibnamefont {Runov}}, \ and\ \bibinfo {author}
  {\bibfnamefont {J.}~\bibnamefont {Geiger}},\ }\href {\doibase
  http://dx.doi.org/10.1063/1.3187907} {\bibfield  {journal} {\bibinfo
  {journal} {Phys. Plasmas}\ }\textbf {\bibinfo {volume} {16}},\ \bibinfo {eid}
  {082303} (\bibinfo {year} {2009})}\BibitemShut {NoStop}%
\bibitem [{Note2()}]{Note2}%
  \BibitemOpen
  \bibinfo {note} {Note that the limit of weak shear connected continuously
  with the zero shear case in Sec.~\ref {small-wt-sec} as localized modes
  become delocalized slab modes; see also \protect \citet
  {candy-waltz-rosenbluth}.}\BibitemShut {Stop}%
\bibitem [{Note3()}]{Note3}%
  \BibitemOpen
  \bibinfo {note} {Note that the toroidal $2\pi /N$-periodicity for
  stellarators formally introduces the possibility of a nonzero Bloch
  wavenumber $k'_{\zeta } = n\pi /N$, with $n$ ranging $0$ to $N -1$, but this
  formal treatment is beyond the scope of the present work.}\BibitemShut
  {Stop}%
\end{thebibliography}%

\end{document}